\documentclass[amsfonts,showpacs,eqsecnum,twocolumn,aps,prd]{revtex4}

\usepackage{amsmath2000}
\usepackage{epsfig}
\usepackage{bbm}

\newcommand\NPA{{Nucl. Phys.} A}
\newcommand\NPB{{Nucl. Phys.} B}

\newcommand\PLB{{Phys. Lett.} B}

\newcommand\PRL{Phys. Rev. Lett.}
\newcommand\PRC{{Phys. Rev.} C}
\newcommand\PRD{{Phys. Rev.} D}

\newcommand\EPJC{{Eur. Phys. J.} C}

\newcommand\JPG{{J. Phys.} G}

\newcommand\CQG{Class. Quant. Grav.}

\newfam\BMath
\font\BMathL=cmmib10 
\font\BMathl=cmmib7
\font\BMathm=cmmib5
\textfont\BMath=\BMathL \scriptfont\BMath=\BMathl
\scriptscriptfont\BMath=\BMathm

\renewcommand\a{\alpha}
\renewcommand\b{\beta}
\newcommand\g{\gamma}
\renewcommand\d{\delta}
\newcommand\e{\epsilon}

\renewcommand\l{\lambda}
\newcommand\m{\mu}
\newcommand\n{\nu}

\newcommand\p{\pi}
\newcommand\s{\sigma}
\renewcommand\t{\tau}

\newcommand\f{\phi}

\renewcommand\j{\psi}
\renewcommand\o{\omega}

\newcommand\D{\Delta}       

\renewcommand\exp{\mbox{\rm exp}}  
\newcommand\tr{\mbox{\rm tr}} 

\newcommand\ra{\rightarrow}

\newcommand\lra{\leftrightarrow}

\newcommand{\lan}{\langle}     
\newcommand{\ran}{\rangle}     

\newcommand\del{\partial}

\newcommand{\fxd}{\!\!}          

\newcommand{\nonum}{\nonumber}

\newcommand{\half}{\frac{1}{2}}
\newcommand{\2}{\frac{1}{2}}

\newcommand{\4}{\frac{1}{4}}

\newcommand\be{\begin{equation}}
\newcommand\ee{\end{equation}}
\newcommand\bea{\begin{eqnarray}}
\newcommand\eea{\end{eqnarray}}
\newcommand\bal{\begin{align}}
\newcommand\eal{\end{align}}
\newcommand\bwt{\begin{widetext}}
\newcommand\ewt{\end{widetext}}

\newcommand\ba{\begin{array}}
\newcommand\ea{\end{array}}
\newcommand\bc{\begin{center}}
\newcommand\ec{\end{center}}

\newcommand\eref[1]{Eq.~(\ref{#1})}
\newcommand\etwref[2]{Eqs.~(\ref{#1}) and (\ref{#2})}

\newcommand\bfi{\begin{figure}}
\newcommand\efi{\end{figure}}
\newcommand\bpi[1]{\begin{picture}#1}
\newcommand\epi{\end{picture}}

\newcommand{\fref}[1]{Fig.~\ref{#1}}

\def\jou#1#2#3#4{{#1} {\bf #2}, #3 (#4)}

\newcommand\sst{\scriptstyle}

\newcommand\thtw{\frac{3}{2}}

\newcommand\rttw{\sqrt{2}}
\newcommand\rtsx{\sqrt{6}}

\newcommand\ve{\varepsilon}
\newcommand\BC{{\mathbb C}}

\newcommand\BZ{{\mathbb Z}}

\newcommand\hapms{\hat a^\pm_s}
\newcommand\hapmt{\hat a^\pm_t}
\newcommand\hamps{\hat a^\mp_s}
\newcommand\hampt{\hat a^\mp_t}

\newcommand\bfa{\text{\boldmath $a$}}
\newcommand\bft{\text{\boldmath $\t$}}

\newcommand\ha{{\hat a}}
\newcommand\hp{{\hat \p}}
\newcommand\hI{{\hat I}}
\newcommand\hJ{{\hat J}}
\newcommand\hL{{\hat L}} 
\newcommand\hP{{\hat \Pi}}

\newcommand\hR{{\hat R}}

\newcommand\shalf{\mbox{$\frac{1}{2}$}}
\newcommand\shlf{{\sst \half}}

\def\bem{\begin{pmatrix}}
\def\eem{\end{pmatrix}}

\newcommand\bse{\begin{subequations}}
\newcommand\ese{\end{subequations}}

\begin{document}

\title{
Equivalence of Classical Skyrmions and Coherent States of Baryons
\\
II. Baryonic Coherent State Construction on Compact Manifolds
}

\author{S.M.H. Wong}
\affiliation{Department of Physics, The Ohio State University, Columbus,
Ohio 43210}

\begin{abstract} 

In connection with the possibility of skyrmion production from small 
domain disoriented chiral condensates formation from heavy ion collisions, 
the direct relation of a classical skyrmion to baryon states is  
examined. It is argued that a skyrmion is a coherent state of baryons. 
The collective coordinate approach of quantization means that the
physical baryon states exist not in flat space but on a compact manifold. 
This requires the construction of coherent states in such a curve space.  
Using the techniques associated with the Segal-Bargmann transform also known
as the coherent state transform used for example in the study of the classical 
limit of quantum gravity, such states can be constructed in the context
of the Skyrme model. They are made up directly of baryon states on $S^3$ but 
with quantum operators on the $SO(3)$ manifold. In terms of wavefunctions,
they are a superposition of the analytic baryon wavefunctions of Adkins, 
Nappi and Witten. The distribution of the baryon states in terms of the
relative probabilities of the baryons inside a skyrmion can therefore be 
determined.  

\end{abstract}

\date{\today}

\null


\maketitle

\section{Introduction}
\label{s:intro}

Recently the possibility of skyrmion formation from disoriented chiral 
condensates (DCC) was raised \cite{kw1,kw2,kw3}. This is in response 
to the discrepancy of data on the rare hyperon yields at the Super Proton 
Synchrotron (SPS) with those from numerical models. The multi-strange 
baryons such as the $\Omega$ and $\bar \Omega$ also exhibit unusual trend 
when compared to the other hadrons \cite{rol,cal,mar,san,wa97}. Comparison
with thermal model shows that the deviations for $\Omega$ and $\Sigma$ were 
found to be concentrated in the low $k_\perp$ region \cite{tr}. All these point 
to an apparently different production mechanism of the multi-strange baryons. 
It may very well be that the mechanism is there also for the non-strange
baryons but they are too abundant for the small fractional increase in the 
yields to be noticeable. Skyrmion production is one such baryon production
mechanism that can provide an explanation for the discrepancy. This channel
of generating baryons through the Kibble mechanism \cite{kib} has been in 
the literature for some time \cite{dg,ek1,ek2,ehk,ks}. But unlike for example
monopole boundary conditions are important for skyrmions and must be properly
taken into account \cite{as}. An important difference between earlier 
considerations and Ref. \cite{kw1,kw2,kw3} is that in the latter skyrmion 
formation is connected to that of DCC and to data at the SPS for the first
time. 

Ever since the first proposal of domain formation during the chiral phase 
transition could lead to observable consequences in the form of fluctuations  
of the charged to total pion ratios \cite{aa,ar,bj1,bk,bkt1,bkt2,rw1}, 
much experimental effort has been spent on the search for DCC. 
No sign of them has been found so far \cite{wa98a,wa98b,wa98c,mmx} at
least not via pions. While this may be due to no DCC formation, 
it could also be that only small domains are able to form. In the latter 
case, observation of DCC through pion distribution becomes impossible  
\cite{rw1,rw2,gm}. This is when the possibility of skyrmion formation
becomes important. Whereas the observation through pions relies   
on having large domain size, the opposite is true for baryon formation
through skyrmion. In the latter case, the smaller the domain size,
the larger the probability of skyrmion formation \cite{ks}. Of course 
there is a natural limit of how small the domain can be. 

In this series of papers, we intend on exploring this possibility and  
try to put this idea on a more concrete foundation than Ref. \cite{kw1,kw2,kw3} 
were able to do. One of the first requirements is to answer the question of
what a skyrmion really is besides being a baryon. Evidently there are many 
different types of baryons. If a skyrmion is produced, what observable baryon
or baryons are actually produced? There are various attempts in the 
past to try to project out known baryons from chiral solitons for studying 
properties of the baryons \cite{ct,gr,bir,fug,ueh}. The chiral solitons
themselves are invariably constructed out of three valence quarks surrounded  
by a cloud of coherent pions. In this type of approach, the coherence is in 
the pion cloud and not in the baryon states. Another distinct approach 
is that of Amado et al \cite{abo,obba} where auxiliary boson fields are used 
to realize the $SU(2) \times SU(2)$ algebra. From these boson operators,
they constructed group theoretic coherent states for any given fixed 
number of bosons. This fixed boson number can be identified with $N_c$ the 
number of colors. Thus by going to the large $N_c$ limit, these group 
theoretic coherent states can be considered to be a skyrmion. In spite
of the success of either of these approaches of being able to decompose the 
solitons into known baryons and therefore provide the answer to the question
of what baryons will be produced by skyrmion formation, we will use a third 
approach already introduced in Ref. \cite{me1,me2}. The chiral soliton approach 
is based on the linear sigma model and relies on the boson fields to be 
linear independent of one another whereas the Skyrme model is a non-linear
sigma model. Therefore the results of the chiral soliton model cannot be
applied to the Skyrme model. The second approach of Amado et al is 
closer to our method but with an important difference. Although both methods 
are based on coherent states, those in Ref. \cite{abo,obba} are of the group 
theoretic type of Perelomov \cite{ap1,ap2} where the states are generated 
by a relevant group action on some fixed state vector. 
In Ref. \cite{anw} Adkins et al have shown that the baryon wavefunctions 
from the collective coordinate approach are all analytic functions on $S^3$. 
Our goal is to obtain coherent states based on these wavefunctions of 
Ref. \cite{anw}. As a consequence the suitable coherent states must be those 
on $S^3$. Those of Amado et al although quite flexible are not the ones 
that we are after. We have no desire to introduce any auxiliary boson fields
and rather be as close to the work of Ref. \cite{anw} as possible. 

The coherent states that we will use are not of Perelomov type but are 
coherent states on compact manifold appropriate to the problem at hand. 
We have outlined the steps in \cite{me1}. These coherent states on a compact 
manifold follows from the generalized Segal-Bargmann transformation or the 
coherent state transform \cite{ha}. Their original application were in
functional analysis \cite{ha} and the classical limit of quantum gravity
\cite{tt0,tt1,kr,krp,hm}. But most of the techniques can be applied to our present
problem. These states constructed by using these techniques possess many of 
the expected properties of coherent states. As discussed in \cite{me1}, 
however, straightforward application on $S^3$ will fail and it turns out
that one must used operators on $SO(3)$ but states on $S^3$. That is
a mixing of ingredient from difference spaces is required. These will be
elaborated. 

The main result of the paper is that a skyrmion can be expressed directly 
in terms of a coherent superposition of baryon states. In the form of 
superposition of wavefunctions, these are the analytic functions on $S^3$ 
derived in \cite{anw}. The probabilities of the distribution of the 
different baryons in the superposition can be written down analytically
in the form of modulus square of wavefunction weighed by an exponential
factor with the exponent given by the ratio of the energy of the particular
state to a fundamental energy scale. Application of the results to actual
DCC formation in heavy ion collisions will be done in a future work \cite{me3}. 

The organization of the paper is as follows. First in Sec. \ref{s:skyrme}
we briefly review the Skyrme model and the results derived in \cite{me2}.
Then we discuss the connection of classical solutions to coherent states
in Sec. \ref{s:cscs}. The main method of constructing coherent states 
on a compact manifold will be introduced in Sec. \ref{s:cs_cm} and applied 
to $SU(2)$ quantized Skyrme model. As mentioned one must overcome an
obstacle in applying the method and this is discussed in Sec. \ref{s:cs-s3}. 
The solution is given next in Sec. \ref{s:so3-op-su2-s}. Finally in
Sec. \ref{s:super}, the superposition of baryon states will be shown and 
examples will be given. A general expression for the probabilities of the 
different states will be written down and will be calculated in some
simple examples.

\section{The Skyrme Model}
\label{s:skyrme}

The Skyrme model has the Lagrangian density \cite{sk1,sk2} 
\be {\cal L}_S = \frac{f_\p^2}{4} \; \tr(\del_\m U\del^\m U^\dagger) 
           +\frac{1}{32g^2}  \; \tr[U^\dagger \del_\m U,U^\dagger \del_\n U]^2
\label{eq:l_s}
\ee 
where 
\be U = \exp \{i \mbox{\boldmath $\t$} \cdot \f /f_\p\} 
      = ( \s + i \mbox{\boldmath $\t$} \cdot \mbox{\boldmath $\p$})/f_\p 
        \; ,
\ee
$f_\p$ is the pion decay constant and $g$ is the $\rho$-$\p$-$\p$ coupling 
\cite{bha,bmss}. This Lagrangian density has the classical skyrmions as the  
solutions to the Euler-Lagrange equation. 

To quantize the Skyrme model around the maximally symmetric classical 
skyrmion solutions 
\be U = U_S = \exp \{i \mbox{\boldmath $\t$}\cdot \mbox{\boldmath $\hat r$}\; 
                     F(r) \}  \;, 
\label{eq:u_s} 
\ee 
time dependence has to be introduced to these static solutions. One can
do this using the well known collective coordinate quantization first done
by \cite{anw} within the context of the Skyrme model. There are two ways to
do this. The well known one is by using $SU(2)$ collective coordinates
\be A = a_0 + i \bfa \cdot \bft \;. 
\label{eq:A2a} 
\ee
Here $A$ is an element of $SU(2)$ where the components are functions of 
time $a_b=a_b(t)$ and subject to the unitary constraint 
\be  a_0^2+\bfa^2 -1 = a_b a_b -1 = 0    \;. 
\label{eq:con} 
\ee
Because of the form of \eref{eq:l_s} given a solution $U_S$,  
\be  U_S' = A\, U_S A^\dagger 
\ee
is also a solution. In terms of $A$ the Lagrangian takes the form 
\be L = 2\l \dot a_b \dot a_b - M  \;.  
\label{eq:sL2} 
\ee  
$M$ is the basic mass scale of the skyrmions and $\l$ has the dimension
of inverse mass. The explicit integral expressions of these constants in
terms of $F(r)$ can be found for example in Ref. \cite{bha,bmss} and is also
given in our previous paper \cite{me2}. The 
conjugate momentum to the coordinate $a_b$ is naturally given by 
$\p_b = 4 \l \dot a_b$. The Hamiltonian is therefore 
\be H = \frac{1}{8\l} \p_b \p_b +M   \;. 
\label{eq:ham}
\ee 
Proper quantization using the Dirac bracket \cite{di,di2} gives
\bse
\label{eq:a-p-com}
\bea [\ha_b, \ha_c] &=& 0                                   \\ {}
     [\ha_b, \hp_c] &=& i (\d_{bc}-\hat a_b \hat a_c)  
\label{eq:a-p-com-2}                                        \\ {} 
     [\hp_b, \hp_c] &=& i (\ha_c \hat \p_b-\ha_b \hat \p_c) 
\eea
\ese
\cite{me2,hkp}. 

Another approach is to map from $SU(2)$ to $SO(3)$ and quantize using
the $SO(3)$ collective coordinates. This approach is known but as far
as we are aware, it has never been done formally or shown explicitly. In the 
literatures, one can at most see a piece of information here and another 
one there. In \cite{me2} we have shown how to do this rigorously using the
Dirac quantization procedure when dealing with a constrained system. 
In this case, one uses the map 
\be  A \t_i A^\dag = \t_j R_{ji}(A)  
\label{eq:su22so3}
\ee
to map an element $A$ of $SU(2)$ to an element $R$ of $SO(3)$. 
The constraints in this case are the requirement of unit determinant 
\be \det R = 1  \;, 
\ee
the identity is also an element of the group and the
inverse is given by the transposed matrix  
\be R R^{-1} = R^{-1} R = {\mathbbm{1}} \;\;\;\;\;\mbox{and}\;\;\;\;\;    
      R^{-1} = R^T                  \;.
\ee 
After some algebraic manipulation, one can arrive at the Lagrangian 
\be  L = \mbox{$\4$} \l\; \dot R_{ij} \dot R_{ij} -M  \; .  
\ee 
With the conjugate momenta to $R_{ij}$ evidently given by 
\be  \Pi_{ij} =  \mbox{$\2$} \l \dot R_{ij}           \; , 
\label{eq:Pi-so3}
\ee
the Hamiltonian follows easily 
\be H =  \mbox{$\frac{1}{\l}$}\; \Pi_{ij} \Pi_{ij} +M   \;. 
\label{eq:ham-so3} 
\ee 
The quantization using $SO(3)$ collective coordinates yield some
unfamiliar but interesting commutators 
\bse
\label{eq:com-so3}
\bea [\hR_{ij}, \hR_{kl}] &=& 0                                           \\ {} 
     [\hR_{ij}, \hP_{kl}] &=& \shalf i (\d_{ik} \d_{jl} 
                             +\e_{ikm} \e_{jln} \hR_{mn}-\hR_{ij} \hR_{kl} )  
                                                                \nonum \\ \\ {}
     [\hP_{ij}, \hP_{kl}]  
    &=&  \mbox{$\4$} i\d_{ik} ( \hR_{ml}\hP_{mj} -\hR_{mj}\hP_{ml} )   \nonum \\ {} 
    & & +\mbox{$\4$} i\d_{jl} ( \hR_{km}\hP_{im} -\hR_{im}\hP_{ki} )   \;.  
\eea
\ese
Details of these derivations have been laid down in \cite{me2}. 

It has been shown in \cite{me2} that in either case, the classical Hamiltonian  
can be expressed universally as 
\be H = \frac{1}{2\l} J^2 + M = \frac{1}{2\l} I^2 + M   
\label{eq:ham-so3-JI}
\ee
independently of whether $SU(2)$ or $SO(3)$ collective coordinates 
were used for the quantization. Quantization in both cases result
in replacing $J$ and $I$ in \eref{eq:ham-so3-JI} by their operator
counterparts. The differential form of the operators are however quite
different \cite{me2}. The same is true for the resulting eigenstates. In 
the case of using $SU(2)$ collective coordinates, quantization produce states 
with both integral and half-integral spin and isospin states. Their
wavefunctions belong to $L^2(S^3)$ and are polynomials in $(a_b+i a_c)$.
The energy of these states depend on the degree of the polynomial $l$ via
\be  E_l = \frac{1}{8\l} l(l+2) + M \;.
\ee 
Only the half-integral spin and isospin states can be interpreted as physical
baryon states. The lowest lying states are shown in Appendix \ref{a:wfn}. 
Quantization with $SO(3)$ collective coordinates, on the other hand, produces
only integral spin and isospin states. None of these are physical particles
\cite{me2}. The wavefunctions are polynomials of products of combinations of
$R_{ij}$. Again the energy of the states depends on the polynomial 
degree $l$ but with a slightly different dependence 
\be  E_l = \mbox{$\frac{1}{2\l}$} l(l+1) + M   \;.
\ee

\section{Classical Solutions and Coherent States}
\label{s:cscs}

Although the static skyrmion solution was used to obtain the Hamiltonian 
in \eref{eq:ham} and \eref{eq:ham-so3}, its only presence is in the two 
constants $\l$ and $M$. The classical solution determines the mass spectrum 
$E_l$ of the quantum states but does not provide a connection of the 
classical solutions to the latter. To find a connection, recall in the case of 
the simple harmonic oscillator (SHO) that Schr\"odinger showed 
a long time ago that the quantum analogs to the classical solutions can be
found \cite{sch}. These are the so-called coherent states which
are superposition of the eigenstates of the quantum system.
Therefore our present problem should be to find coherent states
of baryons which are analogs of the classical skyrmions. If this can be
done then whenever a skyrmion is formed, the quantum analog will provide
information about the distribution of baryons that can be produced via
DCC. The question is how to do this.

There exist in the literature many papers that touched upon coherent 
states, but most of which are for applications far removed from our
problem (see for example Ref. \cite{c-pro}. This just shows how coherent
states are relevant to many different areas of physics. Only those that 
generalize the concept of coherent states outside the context of the 
SHO are potentially relevant. The closest works in the literature are the
chiral soliton model and the interacting boson model treatments already
mentioned in the introduction. There are at least three principal directions 
for generalization: one that is based on the action of a group, another that
is based on minimizing the uncertainty relation and a third is based on
eigenvectors of the annihilation operators. 

The main proponent of the group based method is probably Perelomov
\cite{ap1,ap2} but there are others (for example see the references in
Ref. \cite{kla}). This generalization is based on that aspect of the original
coherent states where they can be considered to be generated by the action
of the Heisenberg-Weyl group on the vacuum vector. That is
\be |\j\ran = D(\a) |0\ran = \exp (\a \hat A^\dagger-\a^* \hat A) |0\ran \;.
\ee
where $\hat A$ and $\hat A^\dagger$ are the annihilation and creation operators
respectively. Apart from an additional trivial phase factor, $D(\a)$ is
a representation of the Heisenberg-Weyl group. The generalization in this
group based method is then
\be |\j\ran = T(g)  |\j_0\ran
\ee
where $g$ is an element of a group $G$ with a representation $T(g)$.
$|\j_0\ran$ is some fixed vector on which the coherent states are
constructed. This generalization is not particularly suited to our
problem since a $|\j_0\ran$ has to be picked which can be quite arbitrary.
Also in the case of $G=SU(2)\times SU(2)$, which is relevant to our problem
because of spin and isospin, the action of the group on the states will only
be restricted to a multiplet of a particular total (iso)spin $j$ if $|\j_0\ran$
is in the $j$ multiplet. If $|\j_0\ran$ on the other hand is already a
superposition of a number of state vectors from different multiplets, the
arbitrariness of $|\j_0\ran$ will be further increased. The fact of having to 
choose a fixed $|\j_0\ran$ at the start already affects to a large degree 
what the coherent state will be. In our problem, this is tantamount to
selecting some baryon states and force the skyrmion identity on
them. This is undesirable and there is little justification for that.

The second approach based on minimizing the uncertainty relation
to $\Delta x \Delta p = \hbar/2$. The resulting coherent states can then
be labeled by the ratio $\Delta x/\Delta p$ since their product is
fixed at the minimum \cite{ns}. However this approach is not terribly
appealing in the sense that the annihilation and creation operators
depend on energy levels. Also why should the alternative,
say $\Delta x \Delta p = \hbar$ or $\Delta x \Delta p = 3\hbar/2$,
would be bad choice for coherent states \cite{kla}. This approach was
proposed to deal with different type of possible potentials. It is not
clear how one can implement this on compact spaces. In the last approach
coherent states are simultaneous eigenstates of the annihilation operators.
As criticized in Ref. \cite{kla}, there is little value from the physical point
of view why such a mathematical criterion alone should automatically give
the resulting coherent states physical meanings.

The above discussion made clear that relying on only one aspect of the
original coherent states and generalizing based on that alone is not a
physically sound procedure. Much stricter criteria must be imposed
which require, at the least, more than one of the above aspects of coherent
states to hold. It should be reasonable to find coherent states
with properties close to or are the generalization of those of the simple
harmonic oscillator. This need not be true in general because there are
many different quantum systems that bear little resemblance to that of
the SHO. Only some features of the original should carry through to these
other systems. In our case the manifold is $S^3$. The Hamiltonian in
\eref{eq:ham} has no potential so classical motions are geodesics or great
circles on the four-dimensional sphere. As is well known that the
one-dimensional projection of this motion onto any of the $a_b$ axes is
SHO-like. The only difference here is the motion in the different dimensions
are not completely independent. One can see the similarity in the
Lagrangian if the constraint \eref{eq:con} is included using the Lagrange
multiplier method
\be L = 2\l \dot a_b \dot a_b -\rho (a_b a_b -1)
        -M  \;,
\ee
where $\rho$ is the Lagrange multiplier. Apart from the constant terms,
this resembles the four-dimensional SHO Lagrangian. It should therefore
be reasonable for the original coherent states of the SHO to be a guide
for the construction of coherent state in later sections.

\section{Constructing Coherent States on a Compact Manifold}
\label{s:cs_cm}

From the discussions in the previous sections, it is evident that coherent
states satisfying the right physical criteria are required for the problem
at hand. Additionally unlike most other situations the coherent states must
be on a compact manifold of $S^3$. Fortunately there is a method to do this. 
For example it has been done on a circle and a two-dimensional sphere by 
Kowalski et al \cite{krp,kr}, and by Hall et al using another approach on 
general $S^n$ \cite{hm}. Kowalski et al based their method on the relation 
between the polar coordinates and the angular momenta, and the analogous 
relation between the conjugate pair $(x,p)$ in the SHO. In the standard case
\be \ha \sim \hat x + i \hat p 
\ee
so we have 
\be \ha |\a \ran = \a |\a \ran 
\ee
and
\be e^{\b \ha} |\a \ran = e^{\b (\hat x+i \hat p)} = e^{\b \a} |\a \ran \;. 
\ee
The analog of this on a circle is 
\be e^{\ha} |\a \ran = e^{\b (\hat \f+i\hat J)} |\a \ran  \;. 
\ee
Kowalski et al method exploited this analogy. Hall et al, on the other hand,
used techniques developed in the studies of analytic functions on
compact manifolds and applied to canonical quantum gravity, namely the
generalized coherent state transform \cite{ha} together with the
complexifier method of Thiemann \cite{tt0}. We will use this second
method below since it is much easier to generalize to any 
dimensions. In spite of the different approaches and from completely different 
points of view, the results of Ref. \cite{krp,kr} turned out to be two special 
cases of those in Ref. \cite{hm}. This strengthens the method and provides
some significance to the so constructed states.

\subsection{The Method}
\label{ss:m}

According to these works, one starts by constructing the annihilation operators.
If $\hat X=(\hat X_1,\hat X_2,\dots)$ are coordinates on a compact manifold 
then the annihilation operator $\hat A=(\hat A_1,\hat A_2,\dots)$ and the 
creation operators $A^\dagger$ are given by
\bea \hat A_i         &=& \hat W \hat X_i \hat W^{-1}
                       =  e^{-\hat C} \hat X_i e^{ \hat C}
\label{eq:a-op}                                            \\
     \hat A_i^\dagger &=& \hat W^{-1} \hat X_i \hat W
                       =  e^{ \hat C} \hat X_i e^{-\hat C}
\label{eq:c-op}
\eea
where $\hat W= e^{-\hat C}$ is there for phase space Wick rotation
(in quantum gravity) with $\hat C$ the complexifier operator being the
generator of this transformation. Given the annihilation operators,
simultaneous eigenstates of these operators can be determined and the 
coherent states will follow from them. 

\eref{eq:a-op} has its origin from the classical version of the phase space 
Wick rotation. Classically the complexifier function $C$ (to be defined below) 
generates the transformation from the real phase space pair $(X,P)$ to the
complex $(X^\BC,P^\BC)$ pair via
\be X^\BC = \sum^\infty_{n=0} \frac{i^n}{n!} \{X, C\}_n
\ee
and similarly for $P^\BC$. Here 
\be \{X,C \}_{n+1} = \{\{X,C\}_n,C\} =
    \underbrace{\{\dots \{\{X,C\}, C\}, \dots ,C\}}_n
\ee
with $\{X,C\}_0 = X$. The latter denotes applying the Poisson bracket $n+1$ 
times. Quantization promotes functions to operators and Poisson brackets to 
commutators
\be  O \ra \hat O \;,\;\;\;\;
     \{O_1, O_2\}  \ra \frac{1}{i} [\hat O_1,\hat O_2] \;.
\ee
Therefore
\be \hat X^\BC = \sum^\infty_{n=0} \frac{1}{n!} [\hat X, \hat C]_n
               = e^{-\hat C} \hat X e^{\hat C}
\label{eq:xc-op}
\ee
with $[\hat X,\hat C]_0=\hat X$ and
$[\hat X,\hat C]_{n+1} = [[\hat X,\hat C]_n, \hat C]$ representing the
repeated application of the commutator $n+1$ times. The complexified
$X$ is the  classical annihilation function $X^\BC=A$. \eref{eq:a-op} is 
therefore the quantized version of the transformation. 

From Ref. \cite{hm} the complexifier function $C$ itself is given by the 
kinetic energy term in the Hamiltonian divided by an energy scale so that
$C$ is dimensionless. This scale defines a fundamental energy scale to the
problem under consideration \cite{tt1}. Introducing $\o$ as this scale
and taking the kinetic energy from a general Hamiltonian decomposed into
kinetic and potential terms
\be  H = T + V  \;,
\ee
the complexifier function is
\be  C = \frac{T}{\o}  \ra \hat C = \frac{\hat T}{\o}
\ee
which becomes the operator $\hat C$ upon quantization.

The annihilation operators are now completely defined. The coherent
states are found by first finding simultaneous eigenstates of 
these operators. On a manifold, one can always define the 
position states $|x\ran$ by 
\be \hat X_i |x\ran = x_i |x\ran
\ee
where $x=(x_1,x_2,\dots )$ represents a point on the manifold.
From the form of \eref{eq:a-op} simultaneous eigenstates of
$\hat A_i$ can be constructed from $|x\ran$. This is done by acting
on $|x\ran$ with the Wick rotator $\hat W$ 
\be |\f\ran = \hat W |x\ran = e^{-\hat C} |x\ran
\ee
so that
\be \hat A_i |\f\ran = \hat A_i \hat W |x\ran = \hat W \hat X_i |x\ran
                     = x_i \hat W |x\ran      = x_i |\f\ran        \;.
\ee
$|\f \ran$ are now simultaneous eigenstates of $\hat A_i$. They are,
however, not yet the coherent states that we are looking for. Remembering
from the SHO, the eigenvalues of the annihilation operators are
in general complex but the $x_i$'s are real. The complex nature
of the eigenvalues contain information both on positions and momenta.
On the other hand eigenvalues $x_i$ being real can carry only half the 
information. To obtain the coherent states from $|\f\ran$, one must 
perform an analytic continuation from real $x$ to complex $x^\BC$. 
Once this last step is taken, the coherent states are
\be   |\j,x^\BC \ran  = \hat W  |x^\BC\ran
\ee
with
\be   \hat A_i |\j,x^\BC \ran = x^\BC_i |\j\ran      \;.
\ee
The eigenvalues in the set $x^\BC=(x^\BC_1,x^\BC_2,\dots)$ serves
also as the label of the coherent states. It follows that there is one
coherent state for every point in phase space, the same as in the SHO. 

Superficially we are following the third direction for generalization of 
the SHO coherent states as discussed in Sec. \ref{s:cscs}. In reality much 
more stringent conditions come automatically with this method. This
has been addressed in a couple of papers by Thiemann et al \cite{tt2,tt3}.

\subsection{An Illustration with the Simple Harmonic Oscillator}
\label{ss:sho}

The last subsection described the method of how to arrive at the coherent
states for a given manifold. Let us illustrate it with a familiar
example of the n-dimensional SHO \cite{tt0,hm}. In this case the
complexifier is $C = P_i^2/2m\o$. From \eref{eq:xc-op} with the usual
commutation relations
\be  [\hat X_i,\hat P_j] = i \d_{ij} \;,\;\;\;\;\;
     [\hat P_i,\hat P_j] = 0 \;,
\ee
the $i$ component of the annihilation operator is 
\be \hat A_i = \hat X_i^\BC = \hat X_i + \frac{i}{m\o} \hat P_i \;.
\label{eq:sho-a}
\ee
Up to a factor of $\sqrt{m\o/2}$ these are the usual annihilation operators.
According to the previous section, a coherent state labeled by
the complex vector $r^\BC =(r^\BC_1,r^\BC_2,\dots,r^\BC_n)$ is
\be  |\j,r^\BC \ran = e^{-\hat P_i^2/2m\o} |r^\BC \ran  \;.
\ee
In this form, it does not resemble a coherent states of the SHO.
To recover the familiar form, one goes to the position representation
\be  \lan x |\j,r^\BC \ran = \lan x |e^{-\hat P_i^2/2m\o} |r^\BC \ran \;.
\ee
Inserting now a complete set of momentum states
\bea \lan x |\j,r^\BC \ran
     &=& \int d^n p \; \lan x |e^{-\hat P_i^2/2m\o}|p\ran\lan p|r^\BC \ran  \nonum \\
     &=& \int \frac{d^n p}{(2\p)^n} \; e^{-p^2/2m\o} e^{ip \cdot (x-r^\BC)} \nonum \\
     &=& \Big (\frac{m\o}{2\p} \Big )^{n/2} \; e^{-m\o (x-r^\BC)^2/2}  \;.
\eea
Writing $r^\BC= r+i \rho$, the wavefunction becomes
\be \lan x |\j,r^\BC \ran = \Big (\frac{m\o}{2\p} \Big )^{n/2}
                            e^{im\o \rho\cdot (x-r)} \;
                            e^{-m\o (x-r)^2/2 +m\o \rho^2/2}  \;.
\ee
After normalization this is the usual coherent state wavefunction
\be \frac{\lan x |\j,r^\BC \ran}{\sqrt{\lan \j,r^\BC|\j,r^\BC\ran}}
     = \Big (\frac{m\o}{\p} \Big )^{n/4} e^{im\o \rho\cdot (x-r)} \;
       e^{-m\o (x-r)^2/2}  \;.
\ee
It has the n-dimensional vector of eigenvalues
\be \sqrt{m\o/2}\,r^\BC_i =\sqrt{m\o/2}\,(r_i+i \rho_i)
\ee
for the correctly normalized annihilation operators. Observe now that 
the introduced fundamental energy scale $\o$ is the fundamental 
frequency (up to $\hbar$) of the oscillator.

In the case of the SHO, the properties of the coherent states are: 
\begin{itemize}  

\item{there is one coherent state per phase space point,}  

\item{they saturate the Heisenberg uncertainty bound,}

\item{the wavefunction peaks at the location given by the label in the 
      classical limit, and}

\item{Ehrenfest theorem is satisfied.} 

\end{itemize}
The first property was seen in the previous section. The remainders have 
been painstakingly shown to hold for general coherent states on compact
manifolds constructed from this method in the classical limit in 
Ref. \cite{tt2,tt3}. The actual proofs are lengthy and involve rather 
heavy mathematics.

\section{Are Coherent States on $S^3$ Skyrmions?}
\label{s:cs-s3}

Returning to the problem at hand, from the Hamiltonian in \eref{eq:ham} 
the complexifier is 
\be C = \frac{1}{8\l \o} \p_b \p_b = \frac{1}{2\l \o} J^2  \;. 
\ee
In terms of the position operators on $S^3$ which are now the
$\hat a_b$, the annihilation operators are
\be \hat A_b = e^{-\frac{1}{2\l \o} \hat J^2} \hat a_b \,
               e^{ \frac{1}{2\l \o} \hat J^2}          \;.
\label{eq:a-op-s3}
\ee
The form of $\hat A_b$ does not resemble \eref{eq:sho-a} at all but it
can be brought into a form much closer in resemblance. This is shown 
in Appendix \ref{a:a-op-su2-alt}. As seen there \eref{eq:a-op-s3} is really the
analog of \eref{eq:sho-a}. In Appendix \ref{a:choice} we address the 
uniqueness issue of $\hat A_b$. 
The corresponding eigenstates labeled by complex variables $a^\BC$ are 
\be |\j,a^\BC \ran = e^{-\frac{1}{2\l \o} \hat J^2} |a^\BC\ran \;.
\label{eq:cs-s3}
\ee
According to the method, this is the general form of the coherent states
on $S^3$ for our Hamiltonian. Mathematically the problem is solved.
Unfortunately on the physical level, we have not reached our goal at all.

\subsection{The Problem with the Coherent States on $S^3$}
\label{ss:prob}

As discussed in Ref. \cite{me1} we would like to expand the skyrmion in terms
of baryon and excited baryon states. \eref{eq:cs-s3} is not yet in this
form. As mentioned in Sec. \ref{s:skyrme} the $SU(2)$ quantized Skyrme 
Hamiltonian admits both integral and half-integral (iso)spin states and not 
all of them are physical. One has to first verify that \eref{eq:cs-s3} carries 
only half-integral (iso)spin states before it can be identified with a classical
skyrmion. In Ref. \cite{me1} it was shown that applying the method on $S^3$
failed to yield a quantum analog to the skyrmion. It fails precisely due
to the presence of unphysical integral (iso)spin states. 

The complete Hilbert space on which the Skyrme Hamiltonian acts is
spanned by $|j,m,n\ran$ where $j,m,n \in \BZ, \BZ+1/2$. Inserting
a complete set of states in \eref{eq:cs-s3} 
\bea |\j,a^\BC \ran &=& \sum_{j,m,n \in \BZ, \BZ+1/2}
     e^{-\frac{1}{2\l \o} \hat J^2} |j,m,n\ran\lan j,m,n|a^\BC\ran    \nonum \\
                    &=& \sum_{j,m,n \in \BZ, \BZ+1/2}
     e^{-\frac{1}{2\l \o} j(j+1)} \j_{jmn}^* (a^\BC) |j,m,n\ran  \;.  \nonum \\
\label{eq:nots}
\eea
Here $\lan j,m,n|a^\BC\ran = \j_{jmn}^* (a^\BC)$ is the complex conjugate
of the wavefunction of the state with quantum numbers ($j,m,n$) evaluated
at the point $a^\BC$ on the complex four-dimensional sphere $S^3_\BC$
(the complex conjugation is performed first for real $a$ before $a=a^\BC$
is set). A complete set for $j=0,1/2,1,3/2$ were given in Ref. \cite{me2}
and only the $j=1/2,3/2$ ones are repeated in Appendix \ref{a:wfn}. 
As discussed in \cite{me2} and in Sec. \ref{s:skyrme} all wavefunctions 
are polynomials made up of $(a_b+ia_c)$ on $S^3$. One can 
therefore analytically continue them to the complex $S^3_\BC$. 
The expansion \eref{eq:nots} shows manifestly the presence of the 
unphysical states. We can conclude that $|\j,a^\BC\ran$ is 
definitely not the classical analog of a skyrmion.

\subsection{Projecting Out the Physical States?}
\label{ss:proj}

\bfi
\epsfig{file=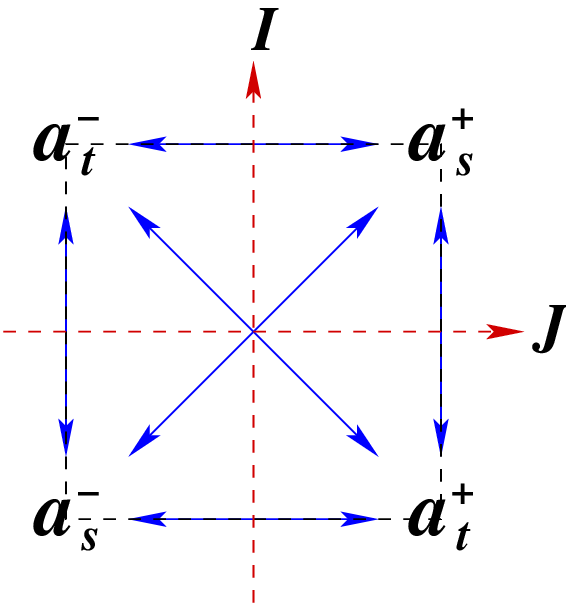,width=4cm}
\caption{This figure summarizes the $\ha^\pm_s$ and $\ha^\pm_t$. Each of
which is a member of simultaneous spin and isospin $j=1/2$ multiplet.
$\hJ_-$ will bring $\ha^+_s$ down to $\ha^-_t$ and $\hI_+$ moves $\ha^-_s$
up to $\ha^-_t$ etc.} 
\label{f:a}
\efi

Can the situation be salvaged somehow? How about simply dropping all
the unphysical states from \eref{eq:nots} or imagine that we have a
projection operator $\hat {\cal P}$ at our disposal, which when applied
to $|\j,a^\BC\ran$, projects out only the physical states
\bea  |\Psi,a^\BC \ran &=& \hat {\cal P} |\j,a^\BC \ran                \nonum \\
      &=& \sum_{j,m,n \in \BZ+1/2} e^{-\frac{1}{2\l \o} j(j+1)}
      |j,m,n\ran\lan j,m,n|a^\BC\ran  \;.                         \nonum \\
\label{eq:cbs?}
\eea
In Ref. \cite{me1} an argument in terms of the wavefunction representation
as to why this would not work was presented. Here we shall instead rely
on the algebra of $S^3$. The form of \eref{eq:a-op} requires us to know 
how $\ha_b$ acts on the states $|j,m,n\ran$. In the Appendix \ref{a:al-s3} 
some algebra of $\hJ_i$ and $\hI_i$ with $\ha_b$ have been worked out.
It is convenient to use the combination 
\be  \hapms = \ha_1 \pm i\ha_2 \;,\;\;\; \hapmt = \ha_0 \pm i\ha_3 \;. 
\ee
The algebra shows that they raise or lower the third component of spin 
and isospin of $|j,m,n\ran$ by $1/2$. For example
\bea \hJ_3 \hapms |j,m,n\ran &=& (m\pm \half) \;\hapms |j,m,n\ran  \\
     \hI_3 \hapmt |j,m,n\ran &=& (n\mp \half) \;\hapmt |j,m,n\ran
\eea
but the $\ha |j,m,n\ran$'s are not eigenstates of $\hJ^2$ and $\hI^2$ in 
general (see \eref{eq:JI2-a-com}). The algebra can succinctly summarized by 
\fref{f:a}. This shows that each of the $\ha^\pm_s$ and $\ha^\pm_t$ belongs
to a simultaneous spin-$1/2$ and isospin-$1/2$ doublets.

Their actions on the state vectors in general are
\bea \hapms |j,m,n\ran &=& C_s^\pm (j,m,n,+) |j+\shlf,m\pm \shlf,n\pm \shlf \ran
                                                         \nonum \\
                       & &+C_s^\pm (j,m,n,-) |j-\shlf,m\pm \shlf,n\pm \shlf \ran
                                                         \nonum \\
                       & & \label{eq:as|st}                     \\
     \hapmt |j,m,n\ran &=& C_t^\pm (j,m,n,+) |j+\shlf,m\pm \shlf,n\mp \shlf \ran
                                                         \nonum \\
                       & &+C_t^\pm (j,m,n,-) |j-\shlf,m\pm \shlf,n\mp \shlf \ran
                                                         \nonum \\
                       & &  \label{eq:at|st}
\eea
where $C_s^\pm$ and $C_t^\pm$ are numerical coefficients and
functions of ($j,m,n$). These equations can be formally deduced from the
algebra of the $SU(2)$ collectively quantized operators. These are given
in Appendix \ref{a:al-s3}. However the simplest way to see this is to use the
observation that each $a_s^\pm$ and $a_t^\pm$ is a member of a doublet.
Therefore they have $j=1/2 $ and $m=n=\pm 1/2$. The eigenstate on which the 
operators act also has its own value of $j,m,n$. The operators and the state 
thus form a direct product which can be decomposed as a direct sum of state 
with new total spin and isospin eigenvalues $j+1/2$ and $j-1/2$
\be j \otimes 1/2 = (j+1/2) \oplus (j-1/2)  \;. 
\ee
This results naturally in \etwref{eq:as|st}{eq:at|st}. In Appendix 
\ref{a:act-ss} more details are given. Also given there are the necessary
steps to solve for the coefficients $C^\pm_s$ and $C^\pm_t$. 

Acting on \eref{eq:cbs?} with $\ha_b$ is to turn every half-integral spin 
state in the expansion into integral spin state. Therefore \eref{eq:cbs?} 
is not an eigenstate of \eref{eq:a-op-s3}. Keeping only physical 
states in the expansion does not solve the problem.

\section{Coherent States with $SO(3)$ Operators on the $S^3$ Manifold}
\label{s:so3-op-su2-s}

We have seen that the problem of constructing coherent state using the 
$SU(2)$ collective quantization and that the $SO(3)$ quantization giving only 
unphysical states. Thus neither approach permits the identification 
of the coherent states as the quantum analog of the classical skyrmions. 
Nevertheless there is a way out. 
As discussed in Ref. \cite{me1} the reason that the $SU(2)$ theory fails
was because of the $SU(2)$ operators mapped fermions to bosons and vice
versa. Therefore the $SU(2)$ annihilation operators require the full
set of bosonic and fermionic states. However if we act on the Hilbert space
of the $SU(2)$ quantized theory exclusively with combinations of $\ha_i \ha_j$, 
this would map fermions to fermions and bosons to bosons. The fermion part and
the boson part of the Hilbert space therefore decouple and the latter 
can be eliminated. This is achieved by ``mixing'' the $SU(2)$ and $SO(3)$
theory. One discards the $SO(3)$ Hilbert space but keeps the operators
and at the same time keeps the Hilbert space (only the fermion half of it)
of the $SU(2)$ theory but discards the operators.

According to the description
in Sec. \ref{ss:m} after introducing an energy scale $\o$, 
from \eref{eq:ham-so3-JI} the complexifier for the $SO(3)$ system is 
\be  C = \frac{1}{\l \o} \Pi_{ij} \Pi_{ij}
       = \frac{1}{2\l \o} J^2               \;.
\ee
The annihilation operators based on the $SO(3)$ coordinate operators $\hR_{ij}$
are therefore
\be \hat A_{ij} = e^{-\frac{1}{2\l \o} \hJ^2}
                  \hR_{ij} e^{\frac{1}{2\l \o} \hJ^2}  \;.
\label{eq:a-op-so3}
\ee
Just like in the $SU(2)$ quantization, the $\hat A_{ij}$ can be expressed
in another form by expanding the exponential and then regrouping the terms. 
This will bring them to a more traditional form. Using the $SO(3)$ algebra
in Appendix \ref{a:al-so3}, this other form is derived in Appendix 
\ref{ss:a-op-so3-alt}. 

The coherent states are constructed from a complex labeled position 
state $|a^\BC\ran$ on $S^3_\BC$ as before. This should be an eigenstate of 
the position operators $\hR_{ij}$. How do the $\hR_{ij}$ act on $|a^\BC\ran$? Recall that 
the $R_{ij}$ although are elements in the $3\times 3$ rotation matrix $R$, 
they are not completely dissociated from the $a_b$ of the $SU(2)$ element $A$. 
In fact as shown in \cite{me2}, $R_{ij} = R_{ij}(A) = R_{ij}(a)$ so
\be  \hR_{ij} |a^\BC \ran = R_{ij}(a^\BC) |a^\BC \ran \;.
\ee 
The position states $|a^\BC\ran$ are indeed eigenstates of $\hR_{ij}$.
Therefore the method together with the above discussion give us 
\be  |\Psi,a^\BC \ran = \hat {\cal P} e^{-\frac{1}{2\l \o} \hJ^2} |a^\BC\ran
\ee
as the coherent states. So with \eref{eq:a-op-so3} as the annihilation 
operators we have 
\be  \hat A_{ij} |\Psi,a^\BC \ran = R_{ij}(a^\BC) |\Psi,a^\BC \ran  \;.
\ee
To obtain these states in terms of the eigenstates $|j,m,n\ran$, one insert 
a ``complete'' set of states in the fermionic space to recover \eref{eq:cbs?}. 
So finally the coherent states consist only of baryon and excited
baryon states that we have been looking for and identifiable with the
classical skyrmions. 

\bfi
\epsfig{file=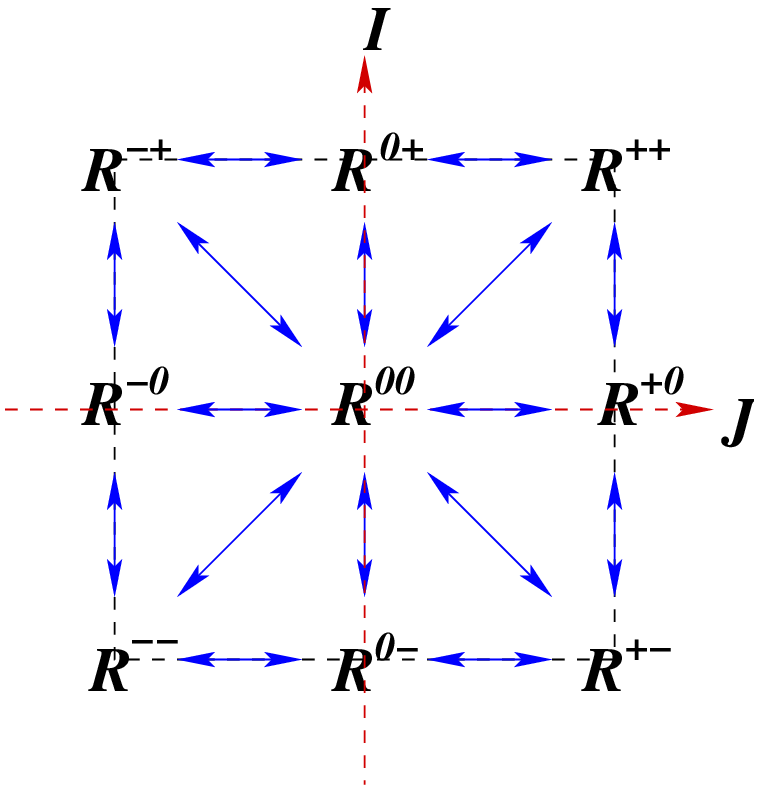,width=5.5cm}
\caption{This figure summarizes at least two important facts.
The first being the action of $\hJ_i$ and $\hI_i$ on the nine combinations
of $\hR_{ij}$. They form three overlapping spin and three isospin triplets.
For example $\hJ^+$ brings $\hR^{00}$ horizontally to $\hR^{+0}$ and $\hI^-$
brings $\hR^{-+}$ vertically down to $\hR^{-0}$ etc (see algebra in Appendix
\eref{eq:JIpm-R-com}. Each member is simultaneously in one spin and one
isospin multiplet. The second is how these combinations act on the states
(see text). For example $\hR^{++}$ raises both the third component of spin
$m$ and isospin $n$ by one, $\hR^{-0}$ lowers $m$ by one but leaves $n$
alone.}
\label{f:R}
\efi

\section{A mixed quantum system of $SO(3)$ operators and $SU(2)$ 
Hilbert Space} 
\label{s:mix-sys}

To permit ourselves to verify explicitly the eigenvalue equations
and to familiarize with this mixed operator-state system, we will derive
the action of the $SO(3)$ operators on the $SU(2)$ eigenstates. 
This can be rigorously deduced from the algebra of the $SO(3)$ operators. 
They are given in the Appendix \ref{a:al-so3}. From the algebra, again one 
can find useful combinations of the components $R_{ij}$. These have been 
given in \cite{me2} and repeated in operator form in Appendix \ref{a:al-so3}. 
Each $R^{pq}$ with $p,q =\pm,0$ is a sum of components $R_{ij}$ and 
belongs simultaneously to one spin and one isospin triplet (see \fref{f:R}). 
Using similar argument as before, the action of $R^{pq}$ on an eigenstate 
$|j,m,n\ran$ results in a direct product between the operator and the state
which can be decomposed as  
\be j \otimes 1 = (j+1) \oplus j \oplus (j-1) \;. 
\ee
Therefore we must have 
\bse
\bea \hR^{pq} |j,m,n\ran &=&\;\; C^{pq}(j,m,n,+) |j+1,m+p1,n+q1\ran     \nonum \\
                         & & +   C^{pq}(j,m,n,0) |j  ,m+p1,n+q1\ran     \nonum \\
                         & & +   C^{pq}(j,m,n,-) |j-1,m+p1,n+q1\ran \;. \nonum \\
\eea
\ese
The $p1,q1$ objects need some explanations. The $p,q$ are the superscript of
$R^{pq}$ and are signs equal to $+$, $-$ or 0. $p1$ means the sign $p$
multiplying with $1$. If $p$ is equal to $+,-$ or $0$, then $p1 =+1$, $-1$
or $0$ respectively. The same applies to $q1$. The coefficients $C^{pq}$
are solved and listed in Appendix \ref{a:solve_coeff} and \ref{a:coeff-so3} 
respectively. The actions of $\Pi^{pq}$ on $|j,m,n\ran$ can be similarly 
expressed but with different coefficients. 

Expectation values are an integral part of any quantum system and it is
most convenient of work in terms of wavefunctions. The 
position wavefunction of the coherent state labeled by $a^\BC$ on
$S^3$ at the point $b=(b_0,b_1,b_2,b_3)$ on $S^3$ would be 
(up to normalization)
\be  \Psi_{a^\BC}(b) = \lan b |\Psi,a^\BC \ran \;. 
\ee
We use $\hR_{ij}$ for position operators so the expectation value of
``position'' would be
\be   \lan \Psi,a^\BC|\hR_{ij}|\Psi,a^\BC \ran 
    = \int d^4 b\; \d(1-b_c^2)\; R_{ij}(b) |\Psi_{a^\BC}(b)|^2  \;.
\ee
This might cause some confusion because the wavefunctions are on $S^3$
while the expectation values are of the $SO(3)$ operators. 
In general expectation values of an operator $\hat O$ from the $SO(3)$
quantized theory in a coherent state is
\be   \lan \Psi,a^\BC|\hat O|\Psi,a^\BC \ran
    = \int d^4 b\; \d(1-b_c^2)\;  \Psi_{a^\BC}(b)^* \,
      \hat O\, \Psi_{a^\BC}(b)  \;.
\ee

\section{Classical Skyrmion as a Coherent Superposition of Baryon and 
Higher Resonance States} 
\label{s:super}

Given that the coherent states are of the form
\be |\Psi,a^\BC \ran = \hat {\cal P} e^{-\frac{1}{2\l \o} \hJ^2} |a^\BC\ran \;,
\ee
inserting a complete set of fermion states and they can  
be written in terms only of baryon states as 
\bea |\Psi,a^\BC \ran &=& \sum_{j,m,n\in \BZ+1/2} e^{-\frac{1}{2\l \o} j(j+1)}
                          |j,m,n\ran \lan j,m,n|a^\BC\ran             \nonum \\  
                      &=& \sum_{j,m,n\in \BZ+1/2} e^{-\frac{1}{2\l \o} j(j+1)}
                          \j^*_{jmn} (a^\BC) |j,m,n\ran  \;.          \nonum \\ 
\eea
$\j_{jmn}(a)$ again are the baryon wavefunctions on $S^3$ 
and $\j^*_{jmn}(a^\BC)$ are the complex conjugates of these 
wavefunctions evaluated at the complex point $a^\BC$.

Let us see some explicit examples of what the states
look like. Using the wavefunctions for $p,n$ ($j=1/2$) and $\Delta$ ($j=3/2$), 
the baryonic coherent state with label $a^\BC=\b=(0,1,0,0)$ is  
\bwt
\bea |\Psi,\b \ran &=& \frac{e^{-\frac{3}{8\l \o}}}{\p} 
  \Big ( |p,{\sst \half}\ran -|n,{\sst-\half}\ran  \Big ) 
                         +\frac{\sqrt{2} e^{-\frac{15}{8\l \o}}}{\p}
  \Big ( |\Delta^{++},{\sst \thtw}\ran -|\Delta^+,{\sst \half}\ran 
        +|\Delta^0,{\sst -\half}\ran -|\Delta^-,{\sst -\thtw}\ran \Big ) 
        +\dots   \;. 
\eea
The ellipses denote half-integral spin higher states. In this case the
nucleons have the same probability and this is similarly true among the 
$\Delta$ in the superposition. The spin and isospin are correlated by 
$m=n$ in this special case. 

With a different label $a^\BC=\g=(\g_0,0,0,\g_3)$ where $\g_0$ and 
$\g_3$ are in general complex numbers satisfying $\g_0^2+\g_3^2=1$, 
the state is  
\bea |\Psi,\g \ran &=& \frac{i e^{-\frac{3}{8\l \o}}}{\p} 
  \Big ( (\g_0+i\g_3) |p,{\sst -\half}\ran  
        -(\g_0-i\g_3) |n,{\sst  \half}\ran \Big )            \nonum \\
  &-&                  \frac{i\sqrt{2} e^{-\frac{15}{8\l \o}}}{\p}  
  \Big ( (\g_0+i\g_3)^3 |\Delta^{++},{\sst -\thtw}\ran   
        -(\g_0+i\g_3)   |\Delta^+   ,{\sst -\half}\ran 
        +(\g_0-i\g_3)   |\Delta^0   ,{\sst  \half}\ran 
        -(\g_0-i\g_3)^3 |\Delta^-   ,{\sst  \thtw}\ran \Big )          \nonum \\   
  &+& \dots   \;. 
\eea
\ewt
Recalling that $\g_i$ are complex so this time both the nucleons and the 
deltas do not have the same probabilities. This is true in general. 
However the special choice of the label $\g$ ensures that this is a
different correlation $m=-n$ in the superposition. 

It should be clear that the expectation values of these states depend 
in general on the label. For example 
\be \lan \Psi,\b|\hJ_i|\Psi,\b\ran = \lan \Psi,\b|\hI_i|\Psi,\b\ran = 0 
\ee 
but
\be \lan \Psi,\g|\hJ_3|\Psi,\g\ran = \lan \Psi,\g|\hI_3|\Psi,\g\ran \ne 0  \;. 
\ee
In the chiral bag soliton type model \cite{ct,gr,bir,fug,ueh}, 
the construction often involves the restriction of ``grand spin'' to 
\be (\hJ_i+\hI_i) |\j\ran = 0    
\ee 
for minimizing the energy of the hedgehog state \cite{fug}. From the last 
expression this can, if need be, be imposed by setting $\g_1=\g_2=\g_3=0$ 
and $\g_0=1$. 

To determine the probability of baryons being produced from skyrmion
in DCC formation, one can calculate the relative probability for a given
baryonic coherent state from the general expression for the states
above. The relative weights between nucleons and Deltas in a state with 
a given label $a^\BC$ is therefore given by  
\be \frac{{\cal P}_\Delta(a^\BC)}{{\cal P}_N(a^\BC)} 
  = e^{-\frac{3}{\l \o}}
    \frac{\sum_{m,n} \j^*_{3/2,m,n} ({a^{\BC}}^*) \j_{3/2,m,n}(a^\BC)}
         {\sum_{m,n} \j^*_{1/2,m,n} ({a^{\BC}}^*) \j_{1/2,m,n}(a^\BC)} \;. 
\ee
The relative probabilities of other higher states can be worked out in
a similar manner. In the first example with $a^\BC=\b$, the relative 
probability between deltas and nucleons is particularly simple 
\be  \frac{{\cal P}_\Delta(\b)}{{\cal P}_N(\b)} 
     = 4\; e^{-\frac{3}{\l \o}} \;. 
\ee
In the second case where $a^\BC =\g$, let us take Im$\g_0=$Re$\g_3 =0$, 
Re$\g_0=\m$ and Im$\g_3=\n =\sqrt{\m^2-1}$ then
\bwt
\be  
     \frac{{\cal P}_\Delta(\g)}{{\cal P}_N(\g)} 
     = e^{-\frac{3}{\l \o}} \left (
       \frac{(\m+\n)^6+(\m-\n)^6+(\m+\n)^2+(\m-\n)^2}{(\m+\n)^2+(\m-\n)^2}
       \right ) \;. 
\ee
\ewt
This shows that in general the expression can be quite complicated but mostly 
one would be interested in the numerical values and these can be worked out 
from the formula provided $\l$ and $\o$ are fixed. The former can be estimated 
from the mass of the nucleon, of the delta and the pion decay constant. 
The latter is not known but a rough guess would be the energy scale
of the chiral phase transition or the value of the temperature $T_c$.

\section{Summary and Outlook}

We have argued that the classical skyrmion solutions can be identified
with their quantum analogs, namely coherent states of baryons. Due to the
non-linear nature of the Skyrme model, the space of the states is the 
curved, compact space of $S^3$. Using the method of Ref. 
\cite{ha,tt0,tt1,kr,krp,hm} specially suitable for compact manifolds, 
these special superposition states of baryons have been successfully 
constructed. The states and wavefunctions are exactly those derived 
by Adkin et al \cite{anw} without modification including the unphysical
states of integral spin and isospin \cite{fr}. Since a skyrmion must be 
made up only of baryons, such states must completely be removed from the 
superposition that ultimately will be equated to the skyrmion. In order 
to overcome this problem, we find that it is necessary to bring in the 
operators from the lesser known $SO(3)$ collective quantization. Only 
with these integral spin and isospin operators can the fermionic and the 
bosonic part of the Hilbert space be decoupled. The unphysical states can 
therefore be discarded. The distribution of the baryon states of given quantum 
numbers in the superposition come in the form of the moduli square of 
the baryon wavefunction of the corresponding states weighed by an 
exponential factor. This factor depends on the ratio of the energy of the 
individual state and an energy scale fundamental to the problem.   
This scale have not been determined but we expect it to be of the
value of the chiral phase transition temperature. With this successful
completion of the decomposition of the skyrmion into know baryon states,
we should be ready to apply this to study skyrmion formation from 
DCC in heavy ion collisions. This will be done in the near future 
\cite{me3}.

\section*{Acknowledgments}
The author thanks K. Kowalski for pointing out Ref. \cite{tt2,hm}, 
B.C. Hall for clarifying the annihilation operators, T. Thiemann for
Ref. \cite{tt1,tt3}, R. Amado and R. Bijker for explaining their papers 
\cite{abo,obba}, P. Ellis, U. Heinz and J. Kapusta for comments. 
This work was supported by the U.S. Department of Energy under grant 
no. DE-FG02-01ER41190.

\appendix

\section{Nucleon and Delta Wavefunctions}  
\label{a:wfn} 

For nucleons, 
\bea 
  \lan a|p, {\sst +\half} \ran \fxd &=&\fxd +\frac{1}{\p} (a_1 + i a_2)      \\
  \lan a|p, {\sst -\half} \ran \fxd &=&\fxd -\frac{i}{\p} (a_0 - i a_3)      \\
  \lan a|n, {\sst +\half} \ran \fxd &=&\fxd +\frac{i}{\p} (a_0 + i a_3)      \\
  \lan a|n, {\sst -\half} \ran \fxd &=&\fxd -\frac{1}{\p} (a_1 - i a_2)      
\eea

For Deltas, 
\bea
  \lan a|\D^{++}, {\sst +\thtw} \ran \fxd &=&\fxd  \frac{\rttw}{\p} (a_1 + i a_2)^3  \\
  \lan a|\D^{+} , {\sst +\thtw} \ran \fxd &=&\fxd i\frac{\rtsx}{\p} (a_1 + i a_2)^2 
                                            (a_0 + i a_3)                     \\
  \lan a|\D^{0} , {\sst +\thtw} \ran \fxd &=&\fxd- \frac{\rtsx}{\p} (a_1 + i a_2) 
                                            (a_0 + i a_3)^2                   \\
  \lan a|\D^{-} , {\sst +\thtw} \ran \fxd &=&\fxd-i\frac{\rttw}{\p} (a_0 + i a_3)^3  
\eea

\bea
  \lan a|\D^{++}, {\sst +\half} \ran \fxd &=&\fxd-i\frac{\rtsx}{\p} (a_1 + i a_2)^2
                                           (a_0 - i a_3)                     \\
  \lan a|\D^{+} , {\sst +\half} \ran \fxd &=&\fxd- \frac{\rttw}{\p} (a_1 + i a_2)
                                           (1 - 3 (a_0^2 + a_3^2))    \nonum \\ \\
  \lan a|\D^{0} , {\sst +\half} \ran \fxd &=&\fxd i\frac{\rttw}{\p} (a_0 + i a_3)
                                           (1 - 3 (a_1^2 + a_2^2))    \nonum \\ \\
  \lan a|\D^{-} , {\sst +\half} \ran \fxd &=&\fxd  \frac{\rtsx}{\p} (a_1 - i a_2)   
                                           (a_0 + i a_3)^2 
\eea

\bea
  \lan a|\D^{++}, {\sst -\half} \ran \fxd &=&\fxd- \frac{\rtsx}{\p} (a_1 + i a_2)
                                           (a_0 - i a_3)^2                   \\
  \lan a|\D^{+} , {\sst -\half} \ran \fxd &=&\fxd-i\frac{\rttw}{\p} (a_0 - i a_3)
                                           (1 - 3 (a_1^2 + a_2^2))    \nonum \\ \\ 
  \lan a|\D^{0} , {\sst -\half} \ran \fxd &=&\fxd  \frac{\rttw}{\p} (a_1 - i a_2)
                                           (1 - 3 (a_0^2 + a_3^2))           \\ 
  \lan a|\D^{-} , {\sst -\half} \ran \fxd &=&\fxd i\frac{\rtsx}{\p} (a_1 - i a_2)^2   
                                           (a_0 + i a_3) 
\eea

\bea
  \lan a|\D^{++}, {\sst -\thtw} \ran \fxd &=&\fxd i\frac{\rttw}{\p} (a_0 - i a_3)^3  \\
  \lan a|\D^{+} , {\sst -\thtw} \ran \fxd &=&\fxd  \frac{\rtsx}{\p} (a_0 - i a_3)^2 
                                            (a_1 - i a_2)                     \\
  \lan a|\D^{0} , {\sst -\thtw} \ran \fxd &=&\fxd-i\frac{\rtsx}{\p} (a_1 - i a_2)^2
                                            (a_0 - i a_3)                     \\
  \lan a|\D^{-} , {\sst -\thtw} \ran \fxd &=&\fxd- \frac{\rttw}{\p} (a_1 - i a_2)^3  
\eea

\section{The Algebra on $S^3$}
\label{a:al-s3}

In the text the action of $J_i$ and $I_i$ on $a_b$ are required. From 
\eref{eq:a-p-com} the commutator between the rotation generators 
(angular momenta) 
\be \hL_{bc} = \ha_b \hp_c - \ha_c \hp_b 
\ee 
and $\ha_b$ can be deduced 
\be  [\hat L_{bc}, \ha_d] = i (\d_{bd} \ha_c - \d_{cd} \ha_b) \;. 
\label{eq:L-a-com}
\ee
It follows that the spin and isospin operators and $\ha_b$ satisfy
\bea   [\hJ_i,\ha_0]  &= \;\;\; \half i \ha_i \;,\;\;
       [\hI_i,\ha_0]  &= -\half i \ha_i                    \\ {}
       [\hJ_i,\ha_i]  &= -\half i \ha_0 \;,\;\;
       [\hI_i,\ha_i]  &= \;\;\; \half i \ha_0              \\ {}
       [\hJ_i,\ha_j]  &=  \half i\e_{ijk} \ha_k  \;,\;\;
       [\hI_i,\ha_j]  &=  \half i\e_{ijk} \ha_k  \;.
\eea
These can be written more compactly as 
\bse
\bea [\hat J_i, \ha_b] &=& \shalf i ( \d_{b0} \ha_i -\d_{ib} \ha_0
                                     +\d_{bj} \e_{ijk} \ha_k )      \\ {}
     [\hat I_i, \ha_b] &=& \shalf i ( \d_{ib} \ha_0 -\d_{b0} \ha_i
                                     +\d_{bj} \e_{ijk} \ha_k )      \;.
\eea
\ese
In this form, they are not particularly useful. Let us write instead
\be  \ha^\pm_s=\ha_1\pm i\ha_2 \;,\;\;\;\;  \ha^\pm_t=\ha_0\pm i\ha_3     \;.
\ee
These commute with each other because the $\ha_b$'s do.
Their action with respect to spin and isospin are dictated and made
clear by the following algebra
\bse
\label{eq:JI-a-com}
\bea [\hJ_\pm,\hapms] &=\;\;0     \;,\;\;\;\;\;\;\;
     [\hI_\pm,\hapms] &= 0               \label{eq:JI-as-0} \\ {}
     [\hJ_\pm,\hamps] &=-i \hapmt  \;,\;\;
     [\hI_\pm,\hamps] &= i \hampt        \label{eq:JI-as}   \\ {}
     [\hJ_\pm,\hapmt] &=\;\;0     \;,\;\;\;\;\;\;\;
     [\hI_\pm,\hampt] &= 0               \label{eq:JI-at-0} \\ {}
     [\hJ_\pm,\hampt] &=\;i\hapms  \;,\;\;\;
     [\hI_\pm,\hapmt] &=-i \hapms        \label{eq:JI-at}   \\ {}
     [\hJ_3,\hapms]   &=\pm\half \hapms \;,\;\;
     [\hI_3,\hapms]   &=\pm\half \hapms  \\ {}
     [\hJ_3,\hapmt]   &=\pm\half \hapmt \;,\;\;
     [\hI_3,\hapmt]   &=\mp\half \hapmt \;.
\eea
\ese
The last two lines show that $\hapms$ and $\hapmt$ act like raising
and lowering operators of one-half instead of the usual one with respect
to the third component of spin and isospin. But states acted on by
$\hapms$ and $\hapmt$ are not eigenstates of $\hJ^2$ and $\hI^2$
in general. The commutators are
\bse
\label{eq:JI2-a-com}
\bea [\hJ^2, \hapms ] &=&     -i\hampt \hJ_\pm \pm \hapms (\hJ_3 \pm \frac{3}{4}) \\ {}
     [\hJ^2, \hapmt ] &=&\;\;\;i\hamps \hJ_\pm \pm \hapmt (\hJ_3 \pm \frac{3}{4}) \\ {}
     [\hI^2, \hapms ] &=&\;\;\;i\hapmt \hI_\pm \pm \hapms (\hI_3 \pm \frac{3}{4}) \\ {}
     [\hI^2, \hapmt ] &=&     -i\hapms \hI_\mp \mp \hapmt (\hI_3 \mp \frac{3}{4}) \;.
\eea
\ese

Similarly one can work out the commutators with the conjugate momenta
$\hp_b$ to find out how they act on the states. Beginning with the
four-dimensional rotation
\be  [\hat L_{bc}, \hp_d] = i (\d_{bd} \hp_c - \d_{cd} \hp_b) \;.
\ee
This is exactly \eref{eq:L-a-com} but with $\ha_b$ replaced by $\hp_b$.
Therefore $\hp_b$ must have the similar commutators with $J_\pm$, $J_3$,
$I_\pm$ and $I_3$ as $\ha_b$.

\section{Action of $\ha_b$ on the Spin and Isospin States}
\label{a:act-ss}

In constructing coherent states on $S^3$, it is necessary to find out
how the operators $\ha_b$ act on the spin and isospin states $|j,m,n\ran$.
These are largely governed by the algebras given in Appendix \ref{a:al-s3}
and the constraint \eref{eq:con}. Naturally the operator form of the
constraint
\be  \ha_b \ha_b = \ha^+_s \ha^-_s + \ha^+_t \ha^-_t = 1
\label{eq:con-a_st}
\ee
is the Casimir operator. The action of $\ha_b$ on the states must take
that into account. It must also constrain the form that this action will 
take. From the algebra, one can see that they either increase or decrease 
the $m$ and $n$ by one-half. The effect on the total (iso)spin $j$ is however 
less clear. Although the action of $\hapms$ and $\hapmt$ on the states give 
eigenstates of $\hJ_3$ and $\hI_3$, these are not eigenstates of $\hJ^2$ 
or $\hI^2$. Lengthy calculation using the algebra on $S^3$ does show 
that the $\ha^\pm$ acting on a state with $j$ gives two states in general: 
one with $j+1/2$ and the other with $j-1/2$. Nevertheless this can also be 
deduced from a much simpler argument. 

From the algebra in Appendix \ref{a:al-s3}, each $\ha_b$ carries $j=1/2$
so decomposition of a direct product of a state with $j$ with $1/2$ gives
the direct sum
\be  j \otimes 1/2 = (j+1/2) \oplus (j-1/2)    \;.
\ee
Alternatively a state of total (iso)spin $j$ has a wavefunction which 
is a polynomial of degree $l = 2j$ in $(a_b+ia_c)$.  
Acting with $\ha^\pm$ on the state wavefunction will
give a polynomial of $l+1 = 2(j+1/2)$. This seems to indicate that a state
with $j$ will become one with $j+1/2$ but this is not the complete story.
Remember that there is the constraint \eref{eq:con-a_st}, it is therefore
always possible to combine $\ha^\pm$ with one $a^\mp$ in the wavefunction
to reduce some terms in the polynomial from degree $l$ to $l-1=2(j-1/2)$.
As a result one must generally have both possibilities of $j+1/2$ and
$j-1/2$. The action of $\ha^\pm$ on the states can thus be written as
\bse
\label{eq:ats-on-s3}
\bea \hapms |j,m,n\ran &=& C_s^\pm (j,m,n,+) |j+\shlf,m\pm \shlf,n\pm \shlf \ran
                                                         \nonum \\
                       & &+C_s^\pm (j,m,n,-) |j-\shlf,m\pm \shlf,n\pm \shlf \ran
                                                         \nonum \\
                       & &                                      \\
     \hapmt |j,m,n\ran &=& C_t^\pm (j,m,n,+) |j+\shlf,m\pm \shlf,n\mp \shlf \ran
                                                         \nonum \\
                       & &+C_t^\pm (j,m,n,-) |j-\shlf,m\pm \shlf,n\mp \shlf \ran
                           \;.                            \nonum \\
                       & &
\eea
\ese
where $C_s^\pm$ and $C_t^\pm$ are numerical functions of ($j,m,n$).

To solve for the coefficients, one uses the algebra in \eref{eq:JI-a-com}.
For example \etwref{eq:JI-as-0}{eq:JI-at-0} acting on a state give relation
connecting $C_s^\pm$ and $C_t^\pm$ to themselves within a (iso)spin multiplet
and \etwref{eq:JI-as}{eq:JI-at} relate $C_s^\pm$ to $C_t^\pm$.
Then from the complex conjugation of matrix elements
\bse
\bea & & \lan j+q\shlf,m\pm \shlf,n\pm \shlf|\hapms|j,m,n\ran^\ast  \nonum \\
     & &\;\;\;\; =\lan j,m,n|\hamps| j+q\shlf,m\pm \shlf,n\pm \shlf \ran
                                                                           \\
     & & \lan j+q\shlf,m\pm \shlf,n\mp \shlf|\hapmt|j,m,n\ran^\ast  \nonum \\
     & &\;\;\;\; =\lan j,m,n|\hampt| j+q\shlf,m\pm \shlf,n\mp \shlf \ran
\eea
\ese
with $q=\pm$ and using \eref{eq:ats-on-s3}, we arrive at the relations
between the coefficient functions
\bse
\label{eq:c_s-c_t}
\bea C_s^\pm(j,m,n,q)^\ast &=& C_s^\mp(j+q\shlf,m\pm \shlf,n\pm \shlf) \\
     C_t^\pm(j,m,n,q)^\ast &=& C_t^\mp(j+q\shlf,m\pm \shlf,n\mp \shlf) \;.
\eea
\ese
These relations together with \eref{eq:con-a_st} allow us to obtain
up to an overall phase factor
\bse
\label{eq:c_st}
\bea
  C^\pm_s(j,m,n,+) &=& \pm  \sqrt{\frac{(j\pm m+1)(j\pm n+1)}
                                       {2(j+1)(2j+1)}}                    \\
  C^\pm_s(j,m,n,-) &=& \mp  \sqrt{\frac{(j\mp m)(j\mp n)}{2j(2j+1)}}      \\
  C^\pm_t(j,m,n,+) &=& \mp i\sqrt{\frac{(j\pm m+1)(j\mp n+1)}
                                       {2(j+1)(2j+1)}}                    \\
  C^\pm_t(j,m,n,-) &=& \mp i\sqrt{\frac{(j\mp m)(j\pm n)}{2j(2j+1)}} \; .
\eea
\ese

\section{Alternative Form of the $SU(2)$ Annihilation Operators}
\label{a:a-op-su2-alt} 

$SU(2)$ operators are not used in the final results for the coherent states. 
They are to be replaced by those from $SO(3)$. We will show, nevertheless, 
in this section and in the next that there is an alternate form of the 
annihilation operators and that which chosen form of the secondary constraint 
among the various available choices discussed in \cite{me2} is used does not 
affect the uniqueness of the annihilation operators. This implies in turn 
that there is no ambiguity as to the uniqueness of the coherent states that 
follow from the operators. 

The alternate form has a much closer resemblance to those of the simple 
harmonic oscillator than the unfamiliar form of position operators sandwiched 
between two exponentials. The complexifier is 
\be  C = \frac{1}{2\l \o} J^2 = \frac{1}{2\l \o} I^2   \;.   
\ee 
The $\o$ is a fundamental energy scale as discussed in the text. 
The spin and isospin operators are related to the rotation generators
$\hL_{bc}$ by 
\bse
\bea  \hJ_i &=& \half (\hL_{0i} + \half \ve_{ijk} \hL_{jk} )   \\  
      \hI_i &=& \half (\hL_{0i} - \half \ve_{ijk} \hL_{jk} )   \;. 
\eea 
\ese 
The annihilation operators are therefore 
\bea \hat A_b &=& e^{-\frac{1}{2\l \o} \hat J^2} \hat a_b \, 
                  e^{ \frac{1}{2\l \o} \hat J^2}                  \\
              &=& \sum^{\infty}_{n=0} \frac{1}{n!} 
                  \Big (\frac{1}{2\l \o} \Big )^n [\ha_b,\hJ^2]_n  \;.   
\eea
with $[\hat A,\hat B]_0=\hat A$ and 
$[\hat A,\hat B]_{n+1} = [[\hat A,\hat B]_n, \hat C]$ representing the 
repeated application of the commutator $n+1$ times as mentioned in the
main text. Let us work out the first few commutators in the sum. We begin by 
using the basic commutator in Appendix \ref{a:al-s3} to deduce that
the angular momentum operators have the commutators  
\be  [\hat L_{bc}, \ha_d] = i (\d_{bd} \ha_c - \d_{cd} \ha_b) \;.
\ee
Then using (see the Appendix of \cite{me2}) 
\be \hJ^2 = \frac{1}{8} \hL_{bc} \hL_{bc}  \;,
\ee
one gets
\bea [\ha_b, \hJ^2] &=& \frac{1}{4} i (\hL_{cb} \ha_c + \ha_c \hL_{cb} ) 
                                                                 \nonum \\ 
                    &=& \frac{1}{2} i \Big (\hL_{cb} -\frac{1}{2}i(d-1) \d_{cb}
                                      \Big ) \ha_c               \nonum \\
                    &=& i ({\cal M} \ha)_b                   \;. 
\eea
The last line has been written as a matrix equation so that \cite{hm}
\be [\ha_b, \hJ^2]_n = i^n ({\cal M}^n \ha)_b   \;. 
\ee
This is true because 
\be [\hJ^2, \hL_{bc}] = 0 \;. 
\ee
The annihilation operators can therefore be expressed in terms of $\mathcal M$
\be \hat A_b = \Big (\exp \{\mbox{$\frac{i}{2\l \o}$} {\cal M}\} \ha \Big )_b \;. 
\ee

Before proceeding further, we will introduce the parameter $\a$ which was 
used in \cite{me2} to distinguish between the three ways that the second 
class constraint derived from \eref{eq:con} can be implemented quantum 
mechanically. In brief one can choose among any of the three 
$\ha_b \hp_b =0$, $\hp_b \ha_b=0$ or $\ha_b \hp_b+\hp_b \ha_b=0$ forms.
This results in the relation
\be \hp_c = \ha_b \hL_{bc} +i \a \ha_c = \hL_{bc} \ha_b +i (\a -d+1) \ha_c 
\label{eq:aL=p}
\ee
between $\hp_b$ and $\hL_{bc}$. The corresponding value of $\a$ is given 
by $\a=0$, $\a=(d-1)$ and $\a=(d-1)/2$ respectively. 

Using \eref{eq:aL=p}, one can work out 
\be ({\cal M} \ha)_b = \half \Big \{\hp_b -i\Big (\a -\half (d-1) \Big )
                           \ha_b \Big \} \;. 
\label{eq:ma} 
\ee 
The appearance of $\hp_b$ here indicates that there is also the need of 
$[\hp_b, \hJ^2]$. The commutator of $\hp_b$ with $\hL$ is similar to that 
with $\ha_b$ 
\be  [\hat L_{bc}, \hp_d] = i (\d_{bd} \hp_c - \d_{cd} \hp_b) 
\ee 
because the $\hL$ are generators of four-dimensional rotations.  
The commutators are 
\bea [\hp_b, \hJ^2] &=& \frac{1}{4}i(\hL_{cb} \hp_c + \hp_c \hL_{cb} ) \nonum \\ {} 
                    &=& \half i \Big (\hL_{cb} -\frac{1}{2}i(d-1) \d_{cb}
                                \Big ) \hp_c                           \nonum \\ {}
                    &=& i ({\cal M} \hp)_b        \;.                  
\eea
The last line can be worked out explicitly using \eref{eq:aL=p} 
\bea ({\cal M} \hp)_b    
       &=& \half ( \hp_b \ha_c \hp_c - \hp_c \hp_c \ha_b 
                  +i\hp_c \ha_c \ha_b ) -\frac{1}{4} i (d+1) \hp_b \nonum \\ {} 
       &=&-\half (\hp_c \hp_c + \a -d+1) \ha_b                     \nonum \\ {}
       & &+\half i \Big (\a-\half (d+1) \Big ) \hp_b          \;. 
\eea 
It is favorable to replace $\hp_c \hp_c$ by $\hJ^2$ which can be achieved
with \eref{eq:aL=p}. Contracting this equation by $\hp_c$ from the left gives 
\be  \hp_c \hp_c = 4 \hJ^2 - \a (\a-d+1) \hat 1   \;.   
\ee
Then 
\bea ({\cal M} \hp)_b 
       &=& -\half \Big (4 \hJ^2 + (1-\a) (\a -d+1) \Big ) \ha_b    \nonum \\ {}
       & & +\half i \Big (\a-\half (d+1) \Big ) \hp_b              \;.    
\label{eq:mp}
\eea
The explicit appearance of $\a$ indicates that the annihilation operators
depend on which of the three choices are chosen as the quantum version of
the secondary constraint. This is the potential non-uniqueness of $\hat A_b$
that we allured to at the beginning. This ambiguity will be
addressed in the next subsection. 

Following \cite{hm} for the purpose of working out $\hat A$, we let 
\begin{equation*} \ha = \bem 1 \\ 0 \eem    \;,\;\;\;\;
                  \hp = \bem 0 \\ 1 \eem    \;,
\end{equation*}
then the corresponding ${\cal M}$ is 
\be {\cal M} = \half \Big ( 
    \begin{smallmatrix}
          -i(\a -\half (d-1))  &  -(4 \hJ^2 + (1-\a) (\a -d+1) )  \\ 
           1                   &  i (\a-\half (d+1) )             \\ 
    \end{smallmatrix}    \Big )  \;. 
\ee
It is advantageous to introduce a shift 
${\cal M} = \tilde {\cal M} -i{\mathbbm 1}/4$.  
\be \hat A_b = \exp \{\mbox{$\frac{1}{8\l \o}$} \} \; 
              (\exp \{\mbox{$\frac{i}{2\l \o}$} \tilde {\cal M}\} \ha )_b  
\ee
where 
\be \tilde {\cal M} = \half \Big ( 
    \begin{smallmatrix}
          -i(\a -\half d)  &  -(4 \hJ^2 + (1-\a) (\a -d+1) )  \\ 
           1               &  i (\a-\half d)                  \\ 
    \end{smallmatrix}    \Big )  \;. 
\ee
This form has the convenient property that on squaring, it yields a diagonal 
matrix 
\be \tilde {\cal M}^2 = -\Big (\hJ^2 +\4 (\frac{d}{2}-1)^2 \hat 1 \Big ) 
         {\mathbbm 1} = -{\cal J}^2 {\mathbbm 1}  \;. 
\label{eq:cj}
\ee 
Therefore this together with \eref{eq:ma} will allow us to work out $\hat A_b$. 

The series expansion of $\hat A_b$ can be most easily calculated by exploiting
the diagonal form of \eref{eq:cj}. Clearly the even power terms are all diagonal
in the column vectors representation of $\ha$ and $\hp$. The sum of even $n$ 
power terms will therefore be a series in ${\cal J}/2\l\o$ multiplying the $n=0$ 
term which is just $\ha_b$
\bea & & \Big (1 -\frac{\tilde {\cal M}^2}{2!(2\l\o)^2} 
               +\frac{\tilde {\cal M}^4}{4!(2\l\o)^4} + \dots \Big)_{bc} \ha_c
                                                                        \nonum \\
     &=& \Big (1 +\frac{{\cal J}^2}{2!(2\l\o)^2} 
               +\frac{{\cal J}^4}{4!(2\l\o)^4} + \dots \Big) \ha_b \nonum \\  
     &=& \cosh \Big (\frac{\cal J}{2\l\o} \Big ) \ha_b   \;.  
\eea
This even power series is the hyperbolic cosine. The sum of odd power 
terms can be re-expressed as an even power series in $\cal M$ by taking
one power of $\cal M$ to act on $\ha_b$ separately 
\bea & &i\Big (\frac{\tilde {\cal M}}{2\l\o} -\frac{\tilde {\cal M}^3}{3!(2\l\o)^3} 
              +\frac{\tilde {\cal M}^5}{5!(2\l\o)^5} + \dots \Big)_{bc} \ha_c
                                                                        \nonum \\
     & &i\Big (\frac{1}{2\l\o} -\frac{\tilde {\cal M}^2}{3!(2\l\o)^3} 
              +\frac{\tilde {\cal M}^4}{5!(2\l\o)^5} + \dots \Big)_{bc} 
         (\tilde {\cal M} \ha )_c                                       \nonum \\
     &=&i\Big (\frac{1}{2\l\o} +\frac{{\cal J}^2}{3!(2\l\o)^3} 
               +\frac{{\cal J}^4}{5!(2\l\o)^5} + \dots \Big)
         (\tilde {\cal M} \ha )_b                                       \nonum \\
     &=&i \frac{1}{\cal J} \sinh \Big (\frac{\cal J}{2\l\o} \Big ) 
         (\tilde {\cal M} \ha )_b    \;. 
\eea
Since 
\be (\tilde {\cal M} \ha)_b = \shalf \p_b -i(\a-\shalf d) \a_b   \;, 
\ee 
therefore finally this is  
\bea \hat A_b &=& e^{\text{$\frac{1}{8\l \o}$}} 
                  \cosh \Big (\frac{{\cal J}}{2\l \o}\Big ) \ha_b   \nonum \\ 
              & &+e^{\text{$\frac{1}{8\l \o}$}} 
                  \frac{1}{2{\cal J}} \big ( \a-\half d \big ) 
                  \sinh \Big (\frac{{\cal J}}{2\l \o}\Big ) \ha_b   \nonum \\
              & &+i e^{\text{$\frac{1}{8\l \o}$}} 
                  \frac{1}{2{\cal J}}   
                  \sinh \Big (\frac{{\cal J}}{2\l \o}\Big ) \hp_b   \;.  
\label{eq:A-su2} 
\eea 
This is the closest
form that the annihilation operators can be reduced to the familiar 
form of the SHO. As mentioned in Ref. \cite{me1}, the $\hJ$ dependent 
coefficients are peculiar to compact spaces.

\section{Are $\hat A_b$ dependent on the Choice of the Secondary 
Constraint?} 
\label{a:choice}

The explicit $\a$ dependence means that this form of $\hat A_b$ is dependent 
on the exact choice of the secondary constraint. Apparently 
this would lead to a problem because it implies that the coherent states 
are also dependent on the choice of $\a$. In fact this is not the case 
in spite of the apparent evidence to the contrary. It is $\hp_b$ which is 
dependent on $\a$ and not $\hat A_b$. This can easily be seen if one 
substitutes for $\hp_b$ in \eref{eq:A-su2}. 

\begin{enumerate}

\item{The First Choice: $\a=0$}

With this choice, the conjugate momenta are related to the angular 
momenta via
\be  \hp_c = \ha_b \hL_{bc} \;. 
\ee
This obviously satisfies the constraint $\ha_b \hp_b =0$. 
Then the annihilation operators are
\bea \hat A_b &=& e^{\text{$\frac{1}{8\l \o}$}} 
                  \cosh \Big (\frac{{\cal J}}{2\l \o}\Big ) \ha_b   \nonum \\ 
              & &-e^{\text{$\frac{1}{8\l \o}$}} 
                  \frac{d}{4{\cal J}} 
                  \sinh \Big (\frac{{\cal J}}{2\l \o}\Big ) \ha_b   \nonum \\
              & &+i e^{\text{$\frac{1}{8\l \o}$}} 
                  \frac{1}{2{\cal J}}   
                  \sinh \Big (\frac{{\cal J}}{2\l \o}\Big ) \hp_b   
\eea 

\item{The Second Choice: $\a=d-1$} 

The conjugate momenta are now given by
\be  \hp_c = \hL_{bc} \ha_b 
           = \ha_b \hL_{bc} +i(d-1) \ha_c  \;. 
\ee
Obviously $\hp_b \ha_b =0$ is satisfied. The form that $\hat A_b$ take
is
\bea \hat A_b &=& e^{\text{$\frac{1}{8\l \o}$}} 
                  \cosh \Big (\frac{{\cal J}}{2\l \o}\Big ) \ha_b   \nonum \\ 
              & &+e^{\text{$\frac{1}{8\l \o}$}} 
                  \frac{1}{2{\cal J}} \big (\half d - 1 \big ) 
                  \sinh \Big (\frac{{\cal J}}{2\l \o}\Big ) \ha_b   \nonum \\
              & &+i e^{\text{$\frac{1}{8\l \o}$}} 
                  \frac{1}{2{\cal J}}   
                  \sinh \Big (\frac{{\cal J}}{2\l \o}\Big ) \hp_b   
\eea

\item{The Third Choice: $\a=(d-1)/2$} 

This choice gives a symmetric expression for the conjugate momenta
\be  \hp_c = \half (\ha_b \hL_{bc} + \hL_{bc} \ha_b) 
           = \ha_b \hL_{bc} +\half i (d-1) \ha_c   \;. 
\ee
The annihilation operators are 
\bea \hat A_b &=& e^{\text{$\frac{1}{8\l \o}$}} 
                  \cosh \Big (\frac{{\cal J}}{2\l \o}\Big ) \ha_b   \nonum \\ 
              & &-e^{\text{$\frac{1}{8\l \o}$}} 
                  \frac{1}{4{\cal J}} 
                  \sinh \Big (\frac{{\cal J}}{2\l \o}\Big ) \ha_b   \nonum \\
              & &+i e^{\text{$\frac{1}{8\l \o}$}} 
                  \frac{1}{2{\cal J}}   
                   \sinh \Big (\frac{{\cal J}}{2\l \o}\Big ) \hp_b   
\eea

\end{enumerate} 

Now rewrite $\p_b$ in each case in terms of $\hL_{bc}$ and $\ha_b$ the
$\hat A_b$ are identically given by 
\bea \hat A_b &=& e^{\text{$\frac{1}{8\l \o}$}} 
                  \cosh \Big (\frac{{\cal J}}{2\l \o}\Big ) \ha_b   \nonum \\ 
              & &-e^{\text{$\frac{1}{8\l \o}$}} 
                  \frac{d}{4{\cal J}} 
                  \sinh \Big (\frac{{\cal J}}{2\l \o}\Big ) \ha_b   \nonum \\
              & &+i e^{\text{$\frac{1}{8\l \o}$}} 
                  \frac{1}{2{\cal J}}   
                  \sinh \Big (\frac{{\cal J}}{2\l \o}\Big ) \ha_b \hL_{bc}  
\eea 
in all cases so the $\a$ dependence cancels out. 
On $S^3$ the angular momenta $\hL_{bc}$ are more fundamental than $\hp_b$.
The formers are independent on the choice $\a$ but not the latter \cite{me2}. 
It follows that the annihilation operators are independent of the choice
of how to implement the second class constraint. It follows that the coherent
states are independent of $\a$.

\section{The Algebra on $SO(3)$}
\label{a:al-so3}

The basic commutators of the spin and isospin operators with the
components of the $SO(3)$ collective coordinates $\hR_{ij}$ are
\bse
\label{eq:JI-R-com}
\bea  [\hJ_i, \hR_{jk}] &=& \;\;\,i\e_{ijm} \hR_{mk}   \\ {}
      [\hI_i, \hR_{jk}] &=& -i \hR_{jm} \e_{imk}
\eea
\ese
After experimenting with the commutators of $\hJ^\pm$, $\hJ_3$, $\hI^\pm$
and $\hI_3$ with $\hR_{ij}$, we find it useful to form the following
\bse
\bea  \hR^{++}      &=& \hR_{11}-\hR_{22} +i(\hR_{12}+\hR_{21})  \\
      \hR^{+-}      &=& \hR_{11}+\hR_{22} -i(\hR_{12}-\hR_{21})  \\
      \hR^{-+}      &=& \hR_{11}+\hR_{22} +i(\hR_{12}-\hR_{21})  \\
      \hR^{--}      &=& \hR_{11}-\hR_{22} -i(\hR_{12}+\hR_{21})  \\
      \hR^{\pm0}    &=& \hR_{13} \pm i \hR_{23}              \\
      \hR^{0\pm}    &=& \hR_{31} \pm i \hR_{32}              \\
      \hR^{00}\;    &=& \hR_{33}  \;.
\eea
\ese
The superscripts are designed with the following algebras in mind
\bse
\label{eq:JI3-R-com}
\begin{align}
     [\hJ_3, \hR^{+\pm} ] &= \;\;\,  \hR^{+\pm} \;, &
     [\hI_3, \hR^{\pm+} ] &= \;\;\,  \hR^{\pm+} \;, \\ {}
     [\hJ_3, \hR^{-\pm} ] &=       - \hR^{-\pm} \;, &
     [\hI_3, \hR^{\pm-} ] &=       - \hR^{\pm-} \;, \\ {}
     [\hJ_3, \hR^{\pm0} ] &=     \pm \hR^{\pm0} \;, &
     [\hI_3, \hR^{0\pm} ] &=     \pm \hR^{0\pm} \;, \\ {}
     [\hJ_3, \hR^{0\pm} ] &=   0                \;, &
     [\hI_3, \hR^{\pm0} ] &=   0                \;, \\ {}
     [\hJ_3, \hR^{00}   ] &=   0                \;, &
     [\hI_3, \hR^{00}   ] &=   0                \;.
\end{align}
\ese
It is clear that $\hR^{++}$, $\hR^{+-}$ and $\hR^{+0}$ raise the third
component of spin by one unit, $\hR^{--}$, $\hR^{-+}$ and $\hR^{-0}$ lower
that by one unit. Also $\hR^{++}$, $\hR^{-+}$ and $\hR^{0+}$ raise the third
component of isospin by one and $\hR^{--}$, $\hR^{+-}$ and $\hR^{0-}$ lower
that by one. $\hR^{\pm0}$ and $\hR^{0\pm}$ leave the third component of
isospin and spin respectively alone. Finally $\hR^{00}$ leaves
the third component of both spin and isospin unchanged. The first and
second superscripts show how the value of $m$ and $n$ respectively 
will be modified. Unlike the $SU(2)$ operators where $\ha_b$ change 
$m$ and $n$ in steps of one-half, the $\hR_{ij}$ change them by one unit
at a time. The half unit increment or decrement is the source of the problem
in $SU(2)$ which effectively couples the boson states with the fermion
states under $\ha_b$. With the $\Delta m = \Delta n = \pm 1$, one should
now be able to cleanly separate the fermionic Hilbert space from the bosonic
part.

The remaining commutation relations in terms of $\hI_{\pm}$ and $\hJ_{\pm}$
are
\bse
\label{eq:JIpm-R-com}
\begin{align}
     [\hJ_+  , \hR^{+\pm}]   &= 0 \;,                 &
     [\hI_+  , \hR^{\pm+}]   &= 0                     \\ {}
     [\hJ_-  , \hR^{-\pm}]   &= 0 \;,                 &
     [\hI_-  , \hR^{\pm-}]   &= 0                     \\ {}
     [\hJ_+  , \hR^{-\pm}]   &=\;\; 2 \hR^{0\pm} \;,  &
     [\hI_+  , \hR^{\pm-}]   &=\;\; 2 \hR^{\pm0}      \\ {}
     [\hJ_-  , \hR^{+\pm}]   &=   - 2 \hR^{0\pm}  \;, &
     [\hI_-  , \hR^{\pm+}]   &=   - 2 \hR^{\pm0}      \\ {}
     [\hJ_+  , \hR^{0\pm}]   &=   - \hR^{+\pm} \;,    &
     [\hI_+  , \hR^{\pm0}]   &=   - \hR^{\pm+}        \\ {}
     [\hJ_-  , \hR^{0\pm}]   &=\;\;\; \hR^{-\pm} \;,  &
     [\hI_-  , \hR^{\pm0}]   &=\;\;\; \hR^{\pm-}      \\ {}
     [\hJ_\pm, \hR^{\pm0}]   &= 0 \;,                 &
     [\hI_\pm, \hR^{0\pm}]   &= 0                     \\ {}
     [\hJ_\mp, \hR^{\pm0}]   &= \mp 2 \hR^{00}   \;,  &
     [\hI_\mp, \hR^{0\pm}]   &= \mp 2 \hR^{00}        \\ {}
     [\hJ_\pm, \hR^{00}  ]   &= \mp \hR^{\pm0}   \;,  &
     [\hI_\pm, \hR^{00}  ]   &= \mp \hR^{0\pm}   \;.
\end{align}
\ese
All the above algebras are summarized in \fref{f:R}. It shows that each
$R^{pq}$ is simultaneously a member of a spin and isospin triplet with
$j=1$. Acting with it on a state with total spin $j$ is equivalent to a
direct product of $j \otimes 1$ which can be decomposed to give the
direct sum
\be  j \otimes 1 = (j+1) \oplus j \oplus (j-1)   
\ee
as mentioned in the text. 
As a result one should obtain in general
\bse
\label{eq:R-on-s3}
\bea \hR^{pq} |j,m,n\ran &=&\;\; C^{pq}(j,m,n,+) |j+1,m+p1,n+q1\ran     \nonum \\
                         & & +   C^{pq}(j,m,n,0) |j  ,m+p1,n+q1\ran     \nonum \\
                         & & +   C^{pq}(j,m,n,-) |j-1,m+p1,n+q1\ran \;. \nonum \\
\eea
\ese

Note that the third component of spin and isospin are modified according
to the specially designed superscripts $p,q$ as mentioned above. The
coefficient functions can be solved. Similar to those of $SU(2)$, the
vanishing commutators in \eref{eq:JIpm-R-com} provide recursion relations
that link $C^{pq}$ within a (iso)spin multiplet. The connections
between $C^{pq}$ and a different $C^{p'q'}$ come from the non-vanishing
commutators in \eref{eq:JIpm-R-com}. These relations severely restrict the
possibilities of the coefficient functions. To actually solve for them,
one has to make use of the constraints which allow the coefficient functions 
to be equated to actual 
numbers and not just to each other, and also similar relations relating
the complex conjugation of one coefficient to another as in \eref{eq:c_s-c_t}
in the $SU(2)$ theory. For example by simply forming matrix elements from
\eref{eq:R-on-s3} using the fact that the Hermitian conjugate on an
$\hR^{pq}$ operator is to change the sign of its superscripts
\be (\hR^{pq})^\dagger = \hR^{-p,-q}
\ee
thus $(\hR^{++})^\dagger = \hR^{--}$, $(\hR^{+0})^\dagger =\hR^{-0}$ etc,
one can easily obtain
\be C^{pq}(j,m,n,r)^* = C^{-p,-q}(j+r,m+p1,n+q1,-r)   \;.
\ee
The details of how to solve for these coefficients are shown in 
Appendix \ref{a:solve_coeff}.

\section{Alternative Form of the $SO(3)$ Annihilation Operators}
\label{ss:a-op-so3-alt}

In the main text the complexifier expressed in terms of the 
(iso)spin operators are the same no matter whether it is the $SU(2)$ or 
$SO(3)$ collective coordinates are used. From the method described there,
the annihilation operators are 
\bea \hat A_{ij} &=& e^{-\frac{1}{2\l \o} \hJ^2} 
                     \hR_{ij} e^{\frac{1}{2\l \o} \hJ^2}                 \\ 
                 &=& \sum^{\infty}_{n=0} \frac{1}{n!} 
                     \Big (\frac{1}{2\l \o} \Big )^n [\hR_{ij},\hJ^2]_n  \;.   
\eea
We will rewrite this in terms of $\hR_{ij}$ and $\hP_{ij}$ to show that
they can be brought into a form closer to the annihilation operators of
the SHO. To work out the series expansion, we use the basic commutators 
in \eref{eq:JI-R-com}. Proceeding in a similar fashion to the
previous $SU(2)$ theory 
\bea [\hR_{ij}, \hJ^2] &=&-2 i \hJ_k \e_{kim} \hR_{mj} -2 \hR_{ij}  \nonum \\ 
                       &=&-2 i (\hJ_k \e_{kim} +i \d_{im}) \hR_{mj} \nonum \\
                       &=& 2 i ({\cal M} \hR )_{ij}   \;.
\eea
We have again introduced a matrix $\cal M$. Because $\hJ^2$ commutes with
any of its individual $\hJ_i$ component, the application of the commutator
$n$ times can be written as 
\be [\hR_{ij}, \hJ^2]_n = (2i)^n ({\cal M}^n \hR)_{ij}   \;. 
\ee
In terms of $\cal M$, the annihilation operators take the form
\be \hat A_{ij} = (\exp \{\mbox{$\frac{i}{\l\o} {\cal M}\}$} \hR )_{ij}  \;. 
\ee
Matrix multiplying $R_{ij}$ once with $\cal M$ gives  
\be ({\cal M} \hR)_{ij} 
    = -(\hR_{il} \hP_{ml} -\hR_{ml} \hP_{il}) \hR_{mj} -i \hR_{ij} 
    =  2 \hP_{ij}   \;. 
\label{eq:MR}
\ee
The last equality is obtained by using 
\be [\hR_{ik}, \hP_{jk}] = [\hR_{ki}, \hP_{kj}] = i \d_{ij}  
\label{eq:RP-com-contr}
\ee 
which follows from \eref{eq:com-so3} and the constraint \cite{me2} 
\be \Pi_{ik} R_{jk} +R_{ik} \Pi_{jk} = 0   \;. 
\label{eq:c8-12}
\ee 

The commutators with the conjugate momenta are more complicated, we need the
commutators 
\be  [\hJ_i, \hP_{jk}] = \;\;\,i\e_{ijm} \hP_{mk}    \;. 
\label{eq:J-Pi-com} 
\ee
Then 
\bea [\hP_{ij}, \hJ^2] &=& -2 i (\hJ_k \e_{kim} +i \d_{im}) \hP_{mj} \nonum \\
                       &=&  2 i ({\cal M} \hP )_{ij}   
\eea  
where 
\bea ({\cal M} \hP)_{ij} 
     &=&-(\hR_{il} \hP_{ml} -\hR_{ml} \hP_{il}) \hP_{mj} -i \hP_{ij}  \nonum \\
     &=&-(\hR_{il} \hP_{ml} -\hP_{il} \hR_{ml}) \hP_{mj}              \nonum \\  
     &=&-\hP_{il} \hP_{ml} \hR_{mj} - \hR_{il} \hP_{ml} \hP_{mj}  \;. 
\label{eq:MPi-nf}
\eea
The constraints \eref{eq:c8-12} is used to obtain the last line. 
Let us now attempt to obtain a relation between the square of the conjugate 
momenta and that of the spin or isospin operator. 
Starting with 
\bea \hJ_i &=& \shalf \e_{ijk} ( \hR_{jl}\hP_{kl} -\hR_{kl}\hP_{jl} ) \nonum \\ 
     \hI_i &=& \shalf \e_{ijk} ( \hR_{lj}\hP_{lk} -\hR_{lk}\hP_{lj} )  
\label{eq:JI-df}
\eea
and using the contracted commutation relations \eref{eq:RP-com-contr} 
and the constraint \eref{eq:c8-12}, we can write 
\bse 
\label{eq:RP-contr}
\bea \hR_{il} \hP_{jl} &=& \shalf (\hJ_m \e_{ijm} +i \d_{ij} )    \\ 
     \hR_{li} \hP_{lj} &=& \shalf (\hI_m \e_{ijm} +i \d_{ij} )    \;.
\eea
\ese 
Contracting one free index of each of these with themselves lead to 
\bea \hR_{il} \hP_{kl} \hR_{kn} \hP_{jn} &=& -\hP_{il} \hP_{jl}       \nonum \\ 
     &=& \mbox{$\4$} (\hJ_m \e_{ikm} +i \d_{ik} ) (\hJ_n \e_{kjn} +i \d_{kj} ) 
                                                                      \nonum \\
     &=& \mbox{$\4$} (\hJ_j \hJ_i -\hJ^2 \d_{ij} +2i \hJ_k \e_{ijk} - \d_{ij}) 
                                                           \;.        \nonum \\ 
\eea
The first equality is possible becauses of \eref{eq:c8-12} and the 
constraint of $R R^{-1} = {\mathbbm{1}}$. It follows that 
\bea -\hP_{il} \hP_{jl} &=& \mbox{$\4$}  (\hJ_i \hJ_j - \hJ^2 \d_{ij}) 
                           +\mbox{$\4$} i(\hJ_k \e_{ijk} +i \d_{ij})  \nonum \\
                        &=& \mbox{$\4$}  (\hJ_i \hJ_j - \hJ^2 \d_{ij}) 
                           -\mbox{$\4$} i {\cal M}_{ij}     \;.    
\label{eq:PiPi-J} 
\eea 
Proceeding similarly for the second equation in \eref{eq:RP-contr},
we also have 
\be -\hP_{li} \hP_{lj}      
    = \mbox{$\4$} (\hI_i \hI_j -\hI^2 \d_{ij} ) 
     +\mbox{$\4$} i (\hI_k \e_{ijk} +i \d_{ij})  \;.  
\label{eq:PiPi-I}
\ee 
Then after contracting the above expressions with $\hR_{mj}$, the first 
term of \eref{eq:MPi-nf} is 
\be  \hP_{il} \hP_{ml} \hR_{mj} 
   =-\mbox{$\4$} (\hJ_i \hJ_m -\hJ^2 \d_{im}) \hR_{mj} +\shalf i \hP_{ij} \;. 
\ee 
To work out the second term of \eref{eq:MPi-nf}, we will look at each term 
in \eref{eq:PiPi-I} and examine the parts in turns.  First there 
is the 
\be  \hR_{im} (\hI_k \e_{mjk} +i \d_{ij}) = 2 \hP_{ij} 
\ee
which can be calculated in a similar fashion to \eref{eq:MR}. Second
one encounters 
\be  \hR_{im} \hI_m 
    = \shalf \e_{mjk} \hR_{im} ( \hR_{lj}\hP_{lk} -\hR_{lk}\hP_{lj} ) \;. 
\ee 
From the primary constraints of unit determinant \cite{me2}, we can write 
\be  \e_{mjk} \hR_{im} \hR_{lj} = \e_{iln} \hR_{nk} 
\ee
then 
\be  \hR_{im} \hI_m = -\e_{inl} R_{nk} \hP_{lk} = -\hJ_i \;. 
\ee 
This is a useful equations linking $\hJ$ and $\hI$ via a contraction 
with the coordinate component $R_{ij}$. This can be converted to the
reciprocal equation  
\be  \hR_{mi} \hJ_m = - \hI_i  \;. 
\ee
Using these equations and the fact that $\hI^2=\hJ^2$, all $\hI_i$ can be 
converted into $\hJ_i$ 
\be   \hR_{im} \hI_m \hI_j = -\hJ_i \hI_j = \hJ_i \hJ_m \hR_{mj} \;. 
\ee 
Therefore the second term becomes
\be \hR_{il} \hP_{ml} \hP_{mj} 
   = -\mbox{$\4$} (\hJ_i \hJ_m -\hJ^2 \d_{im}) \hR_{mj} +\shalf i \hP_{ij} \;,
\ee
so finally
\be ({\cal M} \hP)_{ij} 
   = \shalf (\hJ_i \hJ_m -\hJ^2 \d_{im}) \hR_{mj} -i\hP_{ij} \;.  
\label{eq:MPi} 
\ee

The simplest way to perform the sum of the series in the definition of 
$\hat A_{ij}$ is to suppress the indices of $\hR$ and $\hP$ and let 
\be  \hR  = \bem 1 \\ 0 \eem    \;,\;\;\;\; 
     \hP  = \bem 0 \\ 1 \eem    \;,
\ee
\cite{hm}. Treating them as column vectors allow us to write $\cal M$ 
using \etwref{eq:MR}{eq:MPi} as
\be {\cal M} = \bigg ( 
    \begin{smallmatrix}
        \;\; 0 \;\;    &  \shalf (\hJ \hJ - \hJ^2 \hat 1)   \\ 
        \;\; 2 \;\;    &  \;\;-i\;\;                        \\   
    \end{smallmatrix}    \bigg )  \;. 
\ee
This form is not particularly convenient so a shift by a constant times
the identity will be introduced as before ${\cal M} = \tilde {\cal M}-i/2$
\be \tilde {\cal M} = \bigg (
    \begin{smallmatrix} 
         \;\; \shalf i\;\; &   \shalf (\hJ \hJ - \hJ^2 \hat 1)   \\ 
         \;\; 2 \;\;       &   \;\; -\shalf i \;\;               \\ 
    \end{smallmatrix}    \bigg )  \;. 
\ee
Now the square of this shifted matrix is diagonal 
\be \tilde {\cal M}^2 =-\Big ( (\hJ^2 \hat 1 -\hJ \hJ) +\mbox{$\4$} \hat 1 \Big ) 
                        \mathbbm 1  
                      =-{\cal J}^2 \mathbbm 1  \;. 
\label{eq:calJ}
\ee
We have used the compact notation ${\cal J}^2$ to represent the spin
operators on the left. Because of this property, the 
annihilation operators can be rewritten as 
\be \hat A_{ij} = \exp \{\mbox{$\frac{1}{2\l\o}$}\} 
    \Big (\exp \{\mbox{$\frac{i}{\l\o} \tilde {\cal M}\}$} \hR \Big )_{ij}
\ee
and can be calculated by separating the even and odd power terms as in the
$SU(2)$ theory. The even power terms will sum to 
\bea & & \Big [\Big (1 - \frac{\tilde {\cal M}^2}{2!(\l\o)^2} 
          +\frac{\tilde {\cal M}^4}{4!(\l\o)^2} + \dots \Big ) R \Big ]_{ij}
                                                                  \nonum \\  
     &=& \Big [\Big (1 + \frac{{\cal J}^2}{2!(\l\o)^2}  
          +\frac{{\cal J}^4}{4!(\l\o)^2} + \dots \Big ) R \Big ]_{ij} 
                                                                  \nonum \\  
     &=& 
         \Big [ \cosh \Big ( \frac{{\cal J}}{\l\o} \Big ) R \Big ]_{ij} \;. 
\eea
We have now restored the $SO(3)$ indices which have been suppressed in 
the equations above and used \eref{eq:calJ}. 
The sum of odd power terms can be written as an even power series in
$\tilde {\cal M}$ multiplying $\tilde {\cal M} \hR$ 
\bea & & \Big [\Big ( \frac{i\tilde {\cal M}}{\l\o} 
          -\frac{i\tilde {\cal M}^3}{3!(\l\o)^3} 
          +\frac{i\tilde {\cal M}^5}{5!(\l\o)^5} + \dots \Big ) R \Big ]_{ij} 
                                                                        \nonum \\ 
     &=&i\Big [\Big ( \frac{1}{\l\o} -\frac{\tilde {\cal M}^2 }{3!(\l\o)^3}
          +\frac{\tilde {\cal M}^4 }{5!(\l\o)^5} + \dots \Big )\Big ]_{ik}   
          (\tilde {\cal M} \hR )_{kj}                                   \nonum \\   
     &=& \Big [\frac{1}{\l\o} +\frac{{\cal J}^2}{3!(\l\o)^3} 
          +\frac{{\cal J}^4}{5!(\l\o)^5} + \dots \Big ]_{ik}            
   (2 i\hP- \shalf \hR)_{kj}                          \nonum \\  
     &=& \Big [ \frac{1}{\cal J} \sinh \Big (\frac{{\cal J}}{\l\o} \Big )
               (2 i\hP- \shalf \hR) \Big]_{ij}  \;.
\eea 

Therefore the alternative expression of the annihilation operators are 
\bea \hat A_{ij} &=& e^{\frac{1}{2\l\o}}   
             \cosh \Big ( \frac{{\cal J}}{\l\o} \Big )_{ik} \hR_{kj}   \nonum \\ 
                 & &- e^{\frac{1}{2\l\o}}   
     \Big [ \frac{1}{2\cal J} \sinh \Big (\frac{{\cal J}}{\l\o} \Big ) \Big ]_{ik} 
            \hR_{kj}                                                   \nonum \\
                 & &+ i e^{\frac{1}{2\l\o}}
     \Big [ \frac{2}{\cal J} \sinh \Big (\frac{{\cal J}}{\l\o} \Big ) \Big ]_{ik} 
            \hP_{kj}       \;. 
\eea
The $\cal J$ used here is a shorthand of the operators in \eref{eq:calJ} 
and is only meaningful in the series expansion in ${\cal J}^2$. 
$\hat A_{ij}$ is now in a more familiar and yet different form with
obvious differences in the $\hJ$ dependent coefficients.

\section{Solving for the Coefficients Functions of $\hR_{ij}$ on the States} 
\label{a:solve_coeff}

To solve for the coefficients \eref{eq:R-on-s3}, we use the commutation 
relations between $\hJ_\pm$, $\hI_\pm$ and $\hR^{pq}$ and the 
$SO(3)$ constraints. The commutators give relations between the 
coefficient functions and the constraints allow them to be solved 
up to some arbitrary overall phase factor. 

For example the vanishing commutators provide recursion relations of the
$C^{pq}$ amongst a (iso)spin multiplet. Let us take 
\be [\hJ_+, \hR^{+\pm}] = 0 
\ee
and
\be [\hJ_-, \hR^{+\pm} ] = -2 \hR^{0\pm}   
\ee 
then 
\bwt
\bea \hJ_+ \hR^{+\pm} |j,m,n\ran 
     &=&  \sum_{r=+,0,-} \sqrt{(j+m+r1+2)(j-m+r1-1)} \;  
          C^{+\pm}(j,m,n,r) |j+r1,m+2,n\pm 1 \ran       \nonum \\
     &=& \hR^{+\pm} \hJ_+ |j,m,n\ran                    
      = \sqrt{(j+m+1)(j-m)} \hR^{+\pm} |j,m+1,n\ran     \nonum \\ 
     &=& \sqrt{(j+m+1)(j-m)} \sum_{r=+,0,-} C^{+\pm}(j,m+1,n,r)
         |j+r1,m+2,n\pm 1 \ran   \;. 
\eea
Equating coefficients give
\be  \sqrt{(j+m+r1+2)(j-m+r1-1)} C^{+\pm}(j,m,n,r) 
     = \sqrt{(j+m+1)(j-m)} C^{+\pm}(j,m+1,n,r)    
\label{eq:cm-cm+1}
\ee
for $r=+,0,-$ which leads to 
\bse
\bea C^{+\pm}(j,m+1,n,+) &=& \sqrt{\frac{j+m+3}{j+m+1}} \;
                             C^{+\pm}(j,m,n,+)                 \\ 
     C^{+\pm}(j,m+1,n,0) &=& \sqrt{\frac{(j+m+2)(j-m-1)}{(j+m+1)(j-m)}} \;
                             C^{+\pm}(j,m,n,0)                 \\ 
     C^{+\pm}(j,m+1,n,-) &=& \sqrt{\frac{j-m-2}{j-m}} \; 
                             C^{+\pm}(j,m,n,-)                 
\eea\ese 
\ewt
These relates the spin multiplet of $j$ amongst the coefficients with
the same $p,q$. Because of their recursive nature, one can guess immediately
from these relations that 
\bse
\bea C^{+\pm}(j,m,n,+) &\propto & \sqrt{(j+m+2)(j+m+1)}  \;\;    \\ 
     C^{+\pm}(j,m,n,0) &\propto & \sqrt{(j+m+1)(j-m)}            \\
     C^{+\pm}(j,m,n,-) &\propto & \sqrt{(j-m)(j-m-1)}    \;.  
\eea\ese 
Note that even though $q=\pm$, the expressions on the right do not have any
dependence on the $q$. This dependence must be in the remaining parts. 
From the commutators \cite{me2}, $p$ is associated with the spin $m$ and
$q$ to the isospin $n$. Therefore the $q$ dependence must be with $n$. 
The symmetry of $(\hJ,p) \lra (\hI,q)$ further suggests that 
\bse
\bea  C^{++}(j,m,n,+)  &\propto &  \sqrt{(j+n+2)(j+n+1)} \;      \\ 
      C^{++}(j,m,n,0)  &\propto &  \sqrt{(j+n+1)(j-n)}           \\ 
      C^{++}(j,m,n,-)  &\propto &  \sqrt{(j-n)(j-n-1)}   \;.
\eea\ese
Combining both the dependence on $m$ and $n$, it can be deduced that 
\bwt
\bse 
\label{eq:C+pm}
\bea C^{+\pm}(j,m,n,+) &= & \sqrt{(j+m+2)(j+m+1)(j\pm n+2)(j\pm n+1)} 
                            f(j,p=+,q=\pm,+)     \\
     C^{+\pm}(j,m,n,0) &= & \sqrt{(j+m+1)(j-m)(j\pm n+1)(j\mp n)} 
                            f(j,p=+,q=\pm,0)     \\
     C^{+\pm}(j,m,n,-) &= & \sqrt{(j-m)(j-m-1)(j\mp n)(j\mp n-1)} 
                            f(j,p=+,q=\pm,-)     \;.  
\eea\ese  
The $f$'s are the functions of proportionality which can only depend on $j$,
$p,$ $q$ and $+,0,-$. 
\bea \hJ_- \hR^{+\pm} |j,m,n\ran 
    &=& \sum_{r=+,0,-} \sqrt{(j+m+r1+1)(j-m+r1)} \;  
        C^{+\pm}(j,m,n,r) |j+r1,m,n\pm 1\ran                      \nonum \\  
    &=& \hR^{+\pm} \hJ_- |j,m,n\ran -2 \hR^{0\pm} |j,m,n\ran      \nonum \\ 
    &=& \sum_{r=+,0,-} \Big (\sqrt{(j+m)(j-m+1)} C^{+\pm}(j,m-1,n,r) 
                   -2 C^{0\pm}(j,m,n,r) \Big ) |j+r1,m,n\pm 1\ran \;. 
\eea
Equating coefficients for $r=+,0,-$, we have  
\bea \sqrt{(j+m+r1+1)(j-m+r1)} C^{+\pm}(j,m,n,r) 
     = \sqrt{(j+m)(j-m+1)} C^{+\pm}(j,m-1,n,r) -2 C^{0\pm}(j,m,n,r) \;.  
\eea
Now using \eref{eq:cm-cm+1} this becomes
\be  \Big (r(2j+r1+1) -2 m \Big ) C^{+\pm}(j,m,n,r) 
   = -2 \sqrt{(j+m+r1+1)(j-m+r1)} \;C^{0\pm}(j,m,n,r)  \;. 
\ee
For each each value of $r$ this is 
\bse
\label{eq:c++c0+}
\bea 
  \sqrt{j-m+1} \;C^{+\pm}(j,m,n,+) &=& - \sqrt{j+m+2}        \;C^{0\pm}(j,m,n,+) \\
  m            \;C^{+\pm}(j,m,n,0) &=&   \sqrt{(j+m+1)(j-m)} \;C^{0\pm}(j,m,n,0) \\
  \sqrt{j+m}   \;C^{+\pm}(j,m,n,-) &=&   \sqrt{j-m-1}        \;C^{0\pm}(j,m,n,-) 
\eea\ese 
Therefore the non-vanishing commutators provide relations between
$C^{pq}$ with different pairs of superscript $p,q$. Using the expressions
above for $C^{+\pm}$ 
\bse
\label{eq:C0pm}
\bea C^{0\pm}(j,m,n,+) &=& - \sqrt{(j+m+1)(j-m+1)(j\pm n+2)(j\pm n+1)} f(j,+,\pm,+) \\   
     C^{0\pm}(j,m,n,0) &=& m \sqrt{(j\pm n+1)(j\mp n)} f(j,+,\pm,0)                 \\   
     C^{0\pm}(j,m,n,-) &=&   \sqrt{(j+m)(j-m)(j\mp n)(j\mp n-1)} f(j,+,\pm,-)     \;.  
\eea\ese 
\ewt  
Other similar expressions for $C^{pq}$ can be similarly deduced.

The $f$'s functions can be determined with the help of the constraints. 
For example with $(i,j)=(3,3)$, the constraints impose  
\be \hR_{31}^2 + \hR_{32}^2 + \hR_{33}^2 = 
    \hR^{0-} \hR^{0+} + (\hR^{00})^2     = \hat 1   \;. 
\ee
Sandwiching this between bra and ket of $|j,m,n\ran$ gives
\bea 
  \sum_{r=+,0,-} \Big (|C^{0+} (j,m,n,r)|^2 &+& |C^{00} (j,m,n,r)|^2 \Big ) \nonum \\ 
 =\sum_{r=+,0,-} \Big (|C^{0-} (j,m,n,r)|^2 &+& |C^{00} (j,m,n,r)|^2 \Big ) = 1 
                                                                            \nonum \\   
\label{eq:solv-C}
\eea  
since $(\hR^{0+})^\dagger = \hR^{0-}$ and $[\hR^{0+}, \hR^{0-}]=0$.  
In general this equation contains the squared modulus of six yet-to-be
determined functions. The existence of the recursion relations means
that only one function for each $r$ in each multiplet needs to be 
determined. Because of $j \ge n$, the states 
\[ |j,m,\pm (j+1)\ran  \;,\;\;\;  |j-1,m,\pm (j+1)\ran 
\]
and 
\[ |j-1,m,\pm j\ran  
\]
do not exist, it follows that  
\bea C^{p\pm}(j,m,\pm j,0) &=& C^{p\pm}(j,m,\pm j,-)            \nonum \\ 
                           &=& C^{00}(j,m,\pm j,-)   = 0   \;. 
\eea
Then one can choose the top state of the $j$ multiplet $n=j$ in the 
first line of \eref{eq:solv-C} or the bottom $n=-j$ in the second to get 
\bea &&  |C^{0\pm} (j,m,\pm j,+)|^2 + |C^{00}(j,m,\pm j,+)|^2   \nonum \\
     && +|C^{00}(j,m,\pm j,0)|^2  = 1  \; . 
\eea
This has three functions but not all of them are independent. 
There exists the commutators
\be [\hI_{\mp}, \hR^{0\pm} ] = \mp 2 \hR^{00} 
\ee
which relate $C^{0\pm} (j,m,\pm j,+)$ to $C^{00}(j,m,\pm j,+)$. 
Taking steps similar to those in arriving at \eref{eq:c++c0+}, one
can get 
\be C^{00}(j,m,n,+) = \mp \sqrt{\frac{j\mp n+1}{j\pm n+2}} 
                      \; C^{0\pm}(j,m,n,+) \;.  
\ee
This finally gives us
\be  (2j+3) |C^{00}(j,m,\pm j,+)|^2 +|C^{00}(j,m,\pm j,0)|^2  = 1  \; . 
\ee

Now repeating the same using the transpose of the $(i,j)=(3,3)$ constraint 
equation 
\be \hR_{13}^2 + \hR_{23}^2 + \hR_{33}^2 = 
    \hR^{-0} \hR^{+0} + (\hR^{00})^2     = \hat 1   
\ee
and
\bea C^{00} (j,m,n,+) &=& \mp \sqrt{\frac{j\mp m+1}{j\pm m+2}} 
                              \; C^{\pm 0}(j,m,n,+)             \\
     C^{00} (j,m,n,0) &=& \frac{m}{\sqrt{(j\pm m+1)(j\mp m)}}   
                              \; C^{\pm 0}(j,m,n,0)      \nonum \\  
\eea
to get 
\bea && \frac{2j+3}{j\mp m+1}   |C^{00}(j,m,\pm j,+)|^2      \nonum \\   
     && \;\;\;\;\;\;\; 
       +\frac{(j+1)j\mp m}{m^2} |C^{00}(j,m,\pm j,0)|^2 = 1  \;. 
\eea
With two equations and two unknowns, the modulus of the coefficients are  
\bse
\bea  |C^{00}(j,m,\pm j,0)|    &=& \frac{m}{j+1}                         \\ 
      |C^{\pm 0}(j,m,\pm j,0)| &=& \frac{\sqrt{(j\pm m+1)(j\mp m)}}{j+1} 
\eea\ese 
and 
\bea  |C^{00}(j,m,\pm j,+)|   &=& \frac{1}{j+1} 
                                  \sqrt{\frac{(j+m+1)(j-m+1)}{2j+3}}  \nonum \\ 
                                                                             \\
      |C^{0\pm}(j,m,\pm j,+)| &=& \sqrt{\frac{2(j+m+1)(j-m+1)}{(2j+3)(j+1)}} \;. 
\eea 
These together with \etwref{eq:C+pm}{eq:C0pm} are sufficient for the $f$'s 
functions to be determined up to a phase factor. For example from the last
expression and \eref{eq:C0pm}
\be |f(j,+,\pm,+)| = \frac{1}{(j+1)\sqrt{(2j+3)(2j+1)}} \;. 
\ee
The complete set of coefficient functions with the arbitrary phase factor 
chosen to be unity are listed in the next section.

\section{Coefficients of the action of $SO(3)$ operators on the states}
\label{a:coeff-so3} 

Let us define two common functions of $j$ 
\be  r(j) = \frac{1}{(j+1)\sqrt{(2j+1)(2j+3)}}    
\ee
and
\be  s(j) = \frac{1}{j(j+1)} 
\ee
which will appear in the coefficients below. 
\bwt
\bea
 C^{pq} (j,m,n,+) &=& pq \;r(j)   \sqrt{(j+pm+2)(j+pm+1)(j+qn+2)(j+qn+1)} \\  
 C^{pq} (j,m,n,0) &=& - s(j)      \sqrt{(j+pm+1)(j-pm)(j+qn+1)(j-qn)}     \\
 C^{pq} (j,m,n,-) &=& pq \;r(j-1) \sqrt{(j-pm)(j-pm-1)(j-qn)(j-qn-1)}     
\eea
for $p,q=\pm$. With either $p=0$ or $q=0$ 
\bea 
 C^{0q} (j,m,n,+) &=& -q \;r(j)   \sqrt{(j+m+1)(j-m+1)(j+qn+2)(j+qn+1)} \\  
 C^{p0} (j,m,n,+) &=& -p \;r(j)   \sqrt{(j+pm+2)(j+pm+1)(j+n+1)(j+n-1)} \\
 C^{0q} (j,m,n,0) &=& -m \;s(j)   \sqrt{(j+qn+1)(j-qn)}                 \\ 
 C^{p0} (j,m,n,0) &=& -n \;s(j)   \sqrt{(j+pm+1)(j-pm)}                 \\ 
 C^{0q} (j,m,n,-) &=&  q \;r(j-1) \sqrt{(j+m)(j-m)(j-qn)(j-qn-1)}       \\ 
 C^{p0} (j,m,n,-) &=&  p \;r(j-1) \sqrt{(j-pm)(j-pm-1)(j+n)(j-n)}       
\eea
Then with both $p=q=0$
\bea
 C^{00} (j,m,n,+) &=& r(j)   \sqrt{(j+m+1)(j-m+1)(j+n+1)(j-n+1)} \\  
 C^{00} (j,m,n,0) &=& -m n \; s(j)                               \\ 
 C^{00} (j,m,n,-) &=& r(j-1) \sqrt{(j+m)(j-m)(j+n)(j-n)}         
\eea 
\ewt


\begin{thebibliography}{99}

\bibitem{kw1}J.I. Kapusta and S.M.H. Wong, \jou{\PRL}{86}{4251}{2001}.

\bibitem{kw2}J.I. Kapusta and S.M.H. Wong, \jou{\JPG}{28}{1929}{2002}. 

\bibitem{kw3}J.I. Kapusta and S.M.H. Wong, hep-ph/0201166, to
appear in the proceedings of ICPAQGP-2001, 
Jaipur, India, Nov 2001, Pramana Journal of Physics. 

\bibitem{rol} Roland G et al \jou{\NPA}{638}{91c}{1998}

\bibitem{cal} Caliandro R et al \jou{\JPG}{25}{171}{1999} 

\bibitem{mar} Margetis S et al \jou{\JPG}{25}{189}{1999} 

\bibitem{san} \v S\'andor L et al \jou{\NPA}{661}{481c}{1999}

\bibitem{wa97} Antinori F et al \jou{\EPJC}{14}{633}{2000} 

\bibitem{tr}G. Torrieri and J. Rafelski, \jou{New Jour. Phys.}{3}{12}{2001}.

\bibitem{kib}T.W.B. Kibble, \jou{J. Phys. A}{9}{1387}{1976}.

\bibitem{dg}T.A. DeGrand, \jou{\PRD}{30}{2001}{1984}. 

\bibitem{ek1}J. Ellis and H. Kowalski, \jou{\PLB}{214}{161}{1988}.

\bibitem{ek2}J. Ellis and H. Kowalski, \jou{\NPB}{327}{32}{1989}.

\bibitem{ehk}J. Ellis, U. Heinz, and H. Kowalski, \jou{\PLB}{233}{223}{1989}. 

\bibitem{ks}J.I. Kapusta and A.M. Srivastava, \jou{\PRD}{52}{2977}{1995}.

\bibitem{as}A.M. Srivastava, \jou{\PRD}{43}{1047}{1991}.

\bibitem{aa}A.A. Anselm, \jou{\PLB}{217}{169}{1989}.

\bibitem{ar}A.A. Anselm and M.G. Ryskin, \jou{\PLB}{266}{482}{1991}.

\bibitem{bj1}J.D. Bjorken, ``What lies ahead?'' in Proceedings of the 
Symposium on the SSC: The Project, the progress, the Physics, Corpus Christi, Texas, 
1991, SLAC-PUB-5673 (unpublished). 

\bibitem{bk}J.-P. Blaizot and A. Krzywicki, \jou{\PRD}{46}{246}{1992}.

\bibitem{bkt1}J.D. Bjorken, K.L. Kowalski and C.C. Taylor, ``Baked Alaska''
in proceedings of Les Rencontres de Physique de la Vallee D'Aoste, La Thuile,
Italy, 1993, SLAC-PUB-6109. 

\bibitem{bkt2}J.D. Bjorken, K.L. Kowalski and C.C. Taylor, 
preprint hep-ph/9309235.

\bibitem{rw1}K. Rajagopal and F. Wilczek, \jou{\NPB}{399}{395}{1993}. 

\bibitem{wa98a}M.M. Aggarwal et al, WA98 Collaboration, 
\jou{\PLB}{420}{169}{1998}.

\bibitem{wa98b}T.K. Nayak, WA98 Collaboration, 
\jou{\NPA}{638}{249c}{1998}. 

\bibitem{wa98c}T.K. Nayak, WA98 Collaboration, 
\jou{Pramana Journal of Physics}{57}{285}{2001} 
and nucl-ex/0103007. 

\bibitem{mmx}T.C. Brooks et al, the MiniMax Collaboration, 
\jou{\PRD}{61}{032003}{2000}. 

\bibitem{rw2}K. Rajagopal and F. Wilczek, \jou{\NPB}{404}{577}{1993}. 

\bibitem{gm}S. Gavin and B. M\"uller, \jou{\PLB}{329}{486}{1994}.

\bibitem{ct}A. Chodos and C.B. Thorn, \jou{\PRD}{12}{2733}{1975}.

\bibitem{gr}B. Golli and M. Rosina, \jou{\PLB}{165}{347}{1985}.

\bibitem{bir}M. Birse, \jou{\PRD}{33}{1934}{1986}.

\bibitem{fug}M. Fiolhais, J.N. Urbano and K. Goeke, \jou{\PLB}{150}{253}{1985}.

\bibitem{ueh}M. Uehara, \jou{Prog. Theor. Phys.}{82}{127}{1989}.

\bibitem{abo}R.D. Amado, R. Bijker and M. Oka, \jou{\PRL}{58}{654}{1987}.

\bibitem{obba}M. Oka, R. Bijker, A. Bulgac and R. D. Amado, 
\jou{\PRC}{36}{1727}{1987}. 

\bibitem{me1}S.M.H. Wong, preprint hep-ph/0202250. 

\bibitem{me2}S.M.H. Wong, hep-ph/0207194. 

\bibitem{ap1}A.M. Perelomov, \jou{Commun. Math. Phys.}{26}{222}{1972}. 

\bibitem{ap2}A.M. Perelomov, {\it Generalized Coherent States and Their
Applications} (Springer-Verlag, 1986).

\bibitem{anw}G.S. Adkins, C.R. Nappi and E. Witten, \jou{\NPB}{228}{552}{1983}.

\bibitem{ha}B.C. Hall, \jou{J. Funct. Anal.}{122}{103}{1994}; 

\bibitem{krp}K. Kowalski, J. Rembieli\'nski and L.C. Papaloucas, 
J. Phys. A {\bf 29}, 4149 (1996).  

\bibitem{kr}K. Kowalski and J. Rembieli\'nski, J. Phys. A {\bf 33}, 6035 (2000).  

\bibitem{hm}B.C. Hall and J.J. Mitchell, \jou{J. Math. Phys.}{43}{1211}{2002}. 

\bibitem{tt0}T. Thiemann, \jou{\CQG}{13}{1383}{1996}.  

\bibitem{tt1}H. Sahlmann, T. Thiemann and O. Winkler, \jou{\NPB}{606}{401}{2001}.

\bibitem{me3}S.M.H. Wong, work in progress. 

\bibitem{sk1} T.H.R. Skyrme, {\it Proc. Roy. Soc. A} {\bf 260}, 127 (1961).

\bibitem{sk2} T.H.R. Skyrme, {\it Nucl. Phys.} {\bf 31}, 556 (1962)

\bibitem{bha}R.K. Bhaduri, {\it Models of the Nucleon: From Quark to Soliton}
(Addison-Wesley Publishing, 1988).

\bibitem{bmss}A.P. Balachandran, G. Marmo, B.S. Skagerstam, and A. Stern,
{\it Classical Topology and Quantum States} (World Scientific Publishing, 
Singapore, 1991). 

\bibitem{di}P.A.M. Dirac, {\it Lectures on Quantum Mechanics} 
(Yeshiva University, New York, 1964).

\bibitem{di2}P.A.M. Dirac, \jou{Can. J. Math.}{2}{129}{1950}.

\bibitem{hkp}S.-T. Hong, Y.-W. Kim, Y.-J. Park, \jou{\PRD}{59}{114026}{1999}.

\bibitem{sch}E. Schr\"odinger, \jou{Naturwissenschaften}{14}{664}{1926}. 

\bibitem{c-pro}
{\it Quantization, Coherent States, and
Complex Structures}, edited by J.-P. Antoine et al, (Plenum Press,
New York, 1995).

\bibitem{kla}J.R. Klauder, quant-ph/0110108.

\bibitem{ns}M.M. Nieto and L.M. Simmons Jr., \jou{\PRL}{41}{207}{1978}.

\bibitem{tt2}T. Thiemann and O. Winkler, \jou{\CQG}{18}{2561}{2001}.  

\bibitem{tt3}T. Thiemann and O. Winkler, \jou{\CQG}{18}{4629}{2001}.  

\bibitem{fr}D. Finkelstein and J. Rubinstein, \jou{J. Math. Phys.}{9}{1762}{1968}.

\end{thebibliography}
\end{document}